\documentclass[preprint2]{aastex}

\shorttitle{X-ray Spectra in the Strong Gravity Regime}
\shortauthors{Dov\v{c}iak, Karas \& Yaqoob}

\newcommand{\fekalfa}{{Fe~K$\alpha$} }                                          
\newcommand{\fekbeta}{{Fe~K$\beta$} }

\received{2003 December 23}
\begin{document}

\title{An Extended Scheme for Fitting X-ray Data
with Accretion Disk Spectra in the Strong Gravity Regime$^\dag$}

\author{M.~Dov\v{c}iak,\altaffilmark{\!1,2} V.~Karas,\altaffilmark{\!1,2}
 and T.~Yaqoob\altaffilmark{\,3,4}}
\altaffiltext{1}{Astronomical Institute, Academy of Sciences, 
 Bo\v{c}n\'{\i}~II, CZ-141\,31~Prague, Czech~Republic.}
\altaffiltext{2}{Charles University, Faculty of Mathematics and Physics, 
 CZ-180\,00 Prague, Czech~Republic.}
\altaffiltext{3}{Department of Physics and Astronomy, Johns Hopkins University,
 Baltimore, MD~21218.}
\altaffiltext{4}{Laboratory for High Energy Astrophysics, 
 NASA/\linebreak[1]Goddard Space Flight Center, Greenbelt, MD~20771.\\~\\
$^\dag$~To appear in The Astrophysical Journal, Supplement Series.}

\email{dovciak@mbox.troja.mff.cuni.cz;
 vladimir.karas@cuni.cz;
 yaqoob@skysrv.pha.jhu.edu} 
   
\begin{abstract}
Accreting black holes are believed to emit X-rays which then mediate
information about strong gravity in the vicinity of the emission region. 
We report on a set of new routines for the {\sc{}xspec} package for 
analysing X-ray spectra of black-hole accretion disks. The new
computational tool significantly extends the capabilities of  the
currently available fitting procedures that include the effects of
strong gravity, and allows one to systematically explore the constraints
on more model parameters than previously possible (for example
black-hole angular momentum).  Moreover, axial symmetry of the disk
intrinsic emissivity is not assumed, although it can be imposed to speed
up the computations. The new routines can be used also as a stand-alone
and flexible code with the capability of handling time-resolved spectra
in the regime of strong gravity. We have used the new code to analyse
the mean X-ray  spectrum from the long {\it{}XMM--Newton} 2001 campaign
of the Seyfert~1 galaxy MCG--6-30-15. Consistent with previous 
findings, we obtained a good fit to the broad Fe~K line profile for a
radial line intrinsic emissivity law in the disk which is not a simple
power law, and for near maximal value of black hole angular momentum.
However, equally good fits can be obtained also for small values of the
black hole angular momentum. The code has been developed with the
aim of allowing precise modelling of relativistic effects. Although we
find that current data cannot constrain the parameters of
black-hole/accretion disk system well, the code allows, for a given
source or situation, detailed investigations of what features  of the
data future studies should be focused on in order to achieve the
goal of uniquely isolating the parameters of such systems.
\end{abstract}


\section{Introduction}
There is now strong evidence that the \fekalfa line emission in some
active galactic nuclei (AGN) and some Galactic X-ray binary black-hole
candidates (BHC) originates, at least in part, from an accretion
disk in a strong gravitational field. A lively debate is aimed at
addressing the question of what the spectral line profiles and the
associated continuum can tell us about the  central black-hole, and
whether they can be used to constrain parameters of the accretion disk
in a nearby zone, about ten gravitational radii or less from the
center. For a recent review for AGNs, see Fabian et al.\ (2000),
Reynolds \& Nowak (2003), and references cited therein. For BHCs, see
Miller et al.\ (2002a), McClintock \&  Remillard (2003) and references
therein. In several sources there is indication of \fekalfa line
emission from within the last stable orbit of a Schwarzschild black hole
(e.g.\ in the Seyfert galaxy MCG--6-30-15; see Iwasawa et al.\ 1996;
Fabian et al.\ 2000; Wilms et al.\ 2001; Martocchia, Matt \& Karas
2002a) while in other cases the emission appears to arise farther from
the black hole (e.g.\ in the microquasar GRS 1915+105, see Martocchia et
al.\ 2002b; for AGNs, see Yaqoob \& Padmanabhan 2004 and references
therein). Often, the results from X-ray line spectroscopy are
inconclusive, especially in the case of low spectral resolution data.
For example, a spinning  black hole is allowed but not required by the
line model of the microquasar V4641 Sgr (Miller et al.\ 2002b).  The
debate still remains open, but  there are good prospects for future
X-ray astronomy missions to be able to use the \fekalfa line to probe
the space-time in the vicinity of a black hole, and in particular to
measure the angular momentum, or the spin, associated with the metric.

One may also be able to study the `plunge region'
(about which very little is known), between the event horizon and the 
last stable orbit, and to determine if any appreciable contribution to
the \fekalfa line emission originates from there 
(Reynolds \& Begelman 1997; Krolik \& Hawley 2002). In addition to the Fe~K
lines, there is some evidence for relativistic {\it soft X-ray} emission 
lines due to the Ly$\alpha$ transitions of oxygen, nitrogen, and
carbon (e.g.\ Mason et al.\ 2003), although the observational support
for this interpretation is still controversial (e.g.\ Lee et al.\ 2001).   

With the greatly enhanced spectral resolution and throughput of
future X-ray astronomy missions, the need arises
for realistic theoretical models of the disk emission
and computational tools that are powerful enough to
deal with complex models and to
allow actual fitting of theoretical models to observational
data. It is worth noting that some of the current data
have been used to address the issue of distinguishing
between different space-time metrics around a black hole,
however, the current models available for fitting X-ray data are
subject to various restrictions which we shall elaborate on in the
present paper.

In this paper we describe a generalised scheme and a code which can be
used with the standard X-ray spectral fitting package, {\sc{}xspec}
(Arnaud 1996). We have in mind general relativity models for black-hole
accretion disks. Apart from a better numerical resolution, the principal
innovations compared to currently available schemes are that the new
model allows one to  (i)~fit for the black-hole spin, (ii)~study the
emission from  the plunge region, and (iii)~specify a more general form
of emissivity  as a function of polar coordinates in the disk plane 
(both for the line and for the continuum). Furthermore, it is also 
possible to (iv)~study time variability of the observed signal and 
(v)~compute Stokes parameters of a polarized signal. Items
(i)--(iii) are immediately applicable to current data and modelling, 
while the last two mentioned features are still mainly of theoretical
interest at present. Time-resolved analysis and polarimetry of accretion
disks are directed towards future applications when the necessary
resolution and the ability to do polarimetry are available in X-rays.
Thus our code has the advantage that it can be used with time-resolved
data for reverberation studies of relativistic accretion disks  (Stella
1990; Reynolds et al.\ 1999; Ruszkowski 2000; Goyder \& Lasenby 2004). 
Also polarimetric analysis can be performed, and this will be extremely
useful because it can add very specific information on
strong-gravitational field effects (Connors, Stark \& Piran 1980; Matt,
Fabian \& Ross 1993; Bao et al.\ 1997). Theoretical spectra with
temporal and  polarimetric information can be analysed with  the current
version of our code and such analysis should provide tighter constrains
on future models than is currently possible.

As mentioned above, Fe~K lines have been reported in the X-ray spectra
of numerous AGNs and BHCs.  These sources frequently exhibit remarkable
variability patterns which are still difficult to understand. Among
these puzzling objects, the Seyfert~1 galaxy MCG--6-30-15 is radically 
distinctive with its broad and skewed \fekalfa feature that persists in
observations in spite of substantial variablity of the continuum.
Interpretation of the Fe~K line in terms of  reflection from a
relativistic black-hole disk has been   found to be rather robust, but
it has not been possible to fully explore the model parameter space.  In
particular, constraints on $a$ have only been obtained by fitting
`inverted photon data' which is made from the real data in a
model-dependent way that already assumes a certain value of $a$. The puzzling
stability of this line has been attributed, amongst other things, to
general relativity effects very close to the black hole (Miniutti et
al.\ 2003; Fabian \& Vaughan 2003). It is thus interesting to know if
and how a generalisation of the standard disk-line scheme together with
accurate computation of general relativistic effects  can add new pieces
of information or whether a more substantial  modification of the whole
picture is needed (for further  discussion and references, see e.g.\
Weaver \& Yaqoob 1998; Krolik  1999; Hartnoll \& Blackman 2001; Dumont
et al.\ 2002). Different kinds of such generalisations have been
proposed. Previous papers indicated that details of the model do matter.
However, a strong gravitational field has  almost invariably been
required. In the present paper, we adopt the assumption that the line
and continuum emission are produced by an irradiated disk  near a
rotating black hole, we relax some of the previous assumptions, and we
search for best-fitting parameters of the model. 

In the \S~2 we describe the main features of the new computational 
tool. We list several variants of the code which address different 
problems of fitting X-ray spectra of black hole plus accretion disk
systems.  In \S~3 we use the code to analyse time-averaged data for
MCG--6-30-15 {\it{}XMM--Newton} CCD data as detailed in Fabian et al.\
2002).  We point out that more complex models may not require a large
value of the black-hole  angular momentum, although the current data are
consistent with a maximally  rotating black hole. Finally, in \S~4, we
summarise our results and state our conclusions.

\begin{table*}[tbh]
\caption{Basic features of the new model in comparison with other 
black hole disk-line models.}
\begin{center}
\begin{small}
\tabletypesize{\scriptsize}
\begin{tabular}{lccccl}
\tableline
\multicolumn{1}{c}{Model} 
 & \multicolumn{4}{c}{Effects that are taken into account\rule[-1.5ex]{0mm}{4.5ex}}
 & \multicolumn{1}{c}{Reference}  \\ 
\cline{2-5}
 & \rule[-3.5ex]{0mm}{8ex}\parbox{29mm}{\footnotesize{}Energy shift of photons/Lensing effect}
 & \parbox{18mm}{\footnotesize{}Black hole angular momentum} 
 & \parbox{18mm}{\footnotesize{}Axisymmetry is assumed}
 & \parbox{20mm}{\footnotesize{}Steady source is assumed} &  \\       
\tableline
{\sf{}diskline}   \rule{0mm}{3ex}  &  yes/no~       & $0$~                 & yes          & yes & Fabian et al.\ (1989) \\
{\sf{}laor}                        &  yes/yes       & $0.998$~             & yes          & yes & Laor (1991) \\
{\sf{}kerrspec}                    &  yes/yes       & $\langle0,1\rangle^{\dag}$   & no~          & yes & Martocchia et al.\ (2000) \\
{\sf{}ky}                          &  yes/yes       & $\langle0,1\rangle$           & no$^{\ddag}$ & no  & This paper \\
\tableline
\end{tabular}
\end{small}
\vspace*{2mm}\par{}
{\parbox{0.9\textwidth}{\footnotesize{}$^{\dag}$~The value of dimension-less $a$ parameter 
is kept frozen.\\
$^{\ddag}$~A one-dimensional version is available for the case of an axisymmetric
disk. In this axisymmetric mode, {\sf{}ky} still allows $a$ and other 
relevant parameters to be fitted (in which case the computational speed of {\sf{}ky} is 
then comparable to {\sf{}laor}). The results can be more accurate 
than those obtained with other routines because of the ability to tune the grid
resolution.}}
\label{tab:models}
\end{center}
\end{table*}

\begin{table}[!tbh]
\caption{Basic versions of the model.}
\begin{center}
\begin{footnotesize}
\tabletypesize{\footnotesize}
\begin{tabular}{l@{~}p{22mm}@{~}p{36mm}}
\tableline
Name \rule[-1.5ex]{0mm}{4.5ex} & Type$^{\dag}$~ & Usage \\
\tableline
{\sf{}KYGline} & additive \rule{0mm}{3ex} & Relativistic spectral line from a black hole disk. \\
{\sf{}KYHrefl} & additive  & Compton reflection with an incident power-law (or a broken power-law) continuum. \\
{\sf{}KYLcr} & additive & Lamp-post Compton reflection model. \\
{\sf{}KYSpot} & To be used as an independent code outside {\sc{}xspec} & Time-dependent spectrum of a pattern co-orbiting with the disk. \\
{\sf{}KYConv}  & convolution & Convolution of the relativistic kernel with intrinsic emissivity across the disk. \\
\tableline
\end{tabular}
\end{footnotesize}
\vspace*{2mm}\par{}
{\vspace*{-1mm}\parbox{0.45\textwidth}{\footnotesize{}$^{\dag}$~Different 
model types correspond to {\sc{}xspec} syntax and are 
defined by the way they act in the overall model and form the 
final spectrum. According to the usual convention in {\sc{}xspec}, 
additive models represent individual {\it emission} spectral components 
which may originate e.g.\ in different regions of the source. 
Additive models are simply superposed in the total signal. 
Multiplicative components (e.g.\ {\sf{}hrefl}, discussed in the 
Appendix~\ref{appb}) multiply the current model by an energy-dependent 
factor. Convolution models modify the model in a more non-trivial 
manner. See Arnaud (1996) for details.}}
\label{tab:versions}
\end{center}
\end{table}

\section{Computational Technique}
The new model, {\sf{}ky}, is suited for use with the {\sc{}xspec} 
package (Arnaud 1996). Several  mutations of {\sf{}ky} were developed,
with an eye to specifically provide different applications, and linked 
with a common ray-tracing subroutine, which therefore does not have to
be touched when the intrinsic (local) emissivity  function in the model is
changed. 

When the relativistic line distortions are computed, the new model is
more accurate than the {\sf{}laor} model (Laor 1991) and faster than
{\sf{}kerrspec}  (Martocchia, Karas \& Matt \ 2000). These are the other
two {\sc{}xspec} models with a similar usage (see also Pariev \& Bromley
1998; Gierli\'{n}ski,  Maciolek-Niedzwiecki \& Ebisawa 2001; Schnittman
\& Bertschinger 2003). It is also important
to compare the results from fully independent relativistic codes since
the calculations are sufficiently complex that significant differences
can arise. Our code is more general than the currently available
alternatives in {\sc xspec}, so it can be used to assess the shortcomings
of the available codes. We compared calculated profiles of
the relativistic line in our model with several models available in
literature. In particular, our time-averaged profiles are in perfect 
agreement with those 
shown in Beckwith \& Done (preprint, 2004). Small (but potentially significant) 
distinctions from other models can be explained by omission of light-focusing 
effect in {\sf{}diskline} and by insufficient resolution in the 
{\sf{}laor} model. The discrepancies become more prominent in extreme
situations of large inclination and/or very narrow ring. Furthermore, 
as discussed in the text below, various uncertainties in the form of intrinsic 
emissivity and especially in the darkening law introduce changes in spectrum 
which could not be examined by older models. For more comparisons see 
Appendix~\ref{appa}.

As far as the continuum component is concerned, an approximation was
adopted  in order to account for Compton reflection. To this end, we
employ a more basic approach than the {\sf{}pexrav} model (Magdziarz \&
Zdziarski 1995). Simplicity of the adopted scheme affords a much higher
computation speed than other Compton reflection models in {\sc{}xspec}.
Further details of the method and the calculations for the ray-tracing 
are given in Appendix~\ref{appa}, and details of the Compton reflection
model we use are given in Appendix~\ref{appb} (see also Dov\v{c}iak et
al.\ 2004).

Among its useful features, the {\sf{}ky} model allows one to fit various 
parameters such as black-hole angular momentum ($a$), observer 
inclination angle relative to the disk axis ($\theta_{\rm{}o}$), 
and the size and shape of the emission area on the disk, 
which can be non-axisymmetric (see 
Table~\ref{tab:models} for a summary of the basic features). 
A straightforward modification of a single subroutine
suffices to alter the prescription for the disk emissivity, 
which is specified either by an analytical formula or in a tabular form. 
Our code allows one to change the mesh spacing and resolution for
the (two-dimensional) polar grid that covers the disk plane, as well  
as the energy vector (the output resolution is eventually determined 
by the detector in use when the model is folded through the instrument response). 
Hence, there is sufficient control of the (improved) accuracy and 
computational speed.

Furthermore, {\sf{}ky} can be run as a stand-alone program 
(detached from {\sc{}xspec}). In this mode there is an option 
for time-variable sources such as orbiting spots,
spiral waves or evolving flares (e.g.\ Czerny et al.\ 2004,
who applied a similar approach to compute the predicted {\sf{}rms}
variability in
a specific flare/spot model).
The improved accuracy of the new model has been achieved in several
ways: (i)~photon rays are integrated in Kerr ingoing coordinates
which follow principal photons, (ii)~simultaneous integration of 
the geodesic deviation equations ensures accurate evaluation of the 
lensing effect, and (iii)~non-uniform and rather fine grids
have been carefully selected.

A new and efficient code has been desirable in order
to be able to perform spectral fits with sufficient 
accuracy and to deal with non-axisymmetric
geometry of accretion flows. Also, obscuration effects 
along the line-of-sight need to be taken into account,
as well as the effects of relativity which act on 
photons, as they propagate through curved spacetime
towards a distant observer. A reasonable trade-off must
be achieved between mathematically elegant
approaches (as initiated by de Felice, Nobili \& Calvani 1974, 
and Cunningham 1975) and straightforward
numerical ray-tracing of a sufficiently huge number
of photons (Karas, Vokrouhlick\'y \& Polnarev 1992; 
Bromley, Chen \& Miller 1997). The former method
is less flexible as far as the source geometry
is concerned. On the other hand, the latter method 
is computationally demanding (this problem can be partly
solved by analytical integration of the rays in terms of
elliptical integrals, which is amenable in the Kerr metric, 
but only at cost of substantial complexity of the 
resulting code; cf.\ Martocchia  et al.\ 2000;
\v{C}ade\v{z} et al.\ 2003). Various strategies
and approximations have been developed by different authors 
(Asaoka 1989; Laor 1991; Kojima 1991;
Bao, Hadrava \& {\O}stgaard 1994; Dabrowski \& Lasenby 2001;
Fanton et al.\ 1997; Matt et al.\ 1993a, b;
Semer\'ak, Karas \& de~Felice 1999; Viergutz 1991; Zakharov 1994).
Taking into account the practical experience with various
approaches, we optimised the new routines with respect to rather 
antagonistic requirements on speed, accuracy, generality and flexibility.

Several versions of the routine have been prefabricated for different
types of sources (Table~\ref{tab:versions}): (i)~an intrinsically narrow
line produced by a disk, (ii)~a relativistically blurred 
Compton-reflection continuum including a primary power-law component, 
(iii)~the lamp-post model emissivity (Martocchia et al.\ 2000), and
(iv) time-dependent spectrum of an orbiting or a free-falling spot.
Default parameter values for the line model correspond to those in  the
{\sf{}laor} model, but numerous options have been added. For example, in
the new model one is able to set the emission  inner radius below the
marginally stable orbit, $r_{\rm{}in}<r_{\rm{}ms}(a)$. One can also
allow $a$ to vary independently, in which case the horizon  radius,
$r_{\rm{}h}(a)\equiv1+\sqrt{1-a^2}$, has to be, and indeed is,  updated
at each step of the fit procedure. We thus define emission radii in
terms of their offset from the horizon. Several arguments have been
advocated in favour of having $r_{\rm{}in}{\neq}r_{\rm{}ms}$ for the disk
emission, but this possibility has never been tested rigorously against
observational data. The set of {\sf{}ky-}routines introduced above
provide the tools to explore black-hole disk models and to actually fit
for their key parameters, namely, $a$, $\theta_{\rm{}o}$, and
$r_{\rm{}in}$. Let us remind the reader that, apart from the model
parameters which are subject to fitting, one may also specify the functional
form of intrinsic emissivity of the disk as a function of polar
coordinates and photon energy.

As an example, an time-independent intrinsic emissivity
$I_{\rm{}loc}(E_{\rm{}loc},r)\,\propto\,E_{\rm{}loc}^{-\Gamma_{\rm{}c}}\,
r^{-\alpha_{\rm{}c}}$ was convolved
with  our ray-tracing routine to produce a power-law type continuum
component with relativistic effects. Compton reflection was taken into
account for intrinsic emissivity in the same approximation as adopted in
the (non-relativistic) model {\sf{}hrefl}. The spectrum of the resulting
relativistic model, {\sf{}kyhrefl}, therefore resembles the outcome of a
multiplicative model, {\sf{}hrefl*powerlaw} in {\sc{}xspec}, which has
been blurred by the relativistic disk kernel. It is worth noting
that the effect of general relativity is practically negligible in the
case of a locally power-law continuum because of the overall smearing of the 
spectrum across the disk surface. A non-trivial part in {\sf{}kyhrefl}
therefore concerns the Compton reflection component and its mutual
normalization with respect to the primary powerlaw originating at
different radii in the disk. 

We have also produced a convolution-type
model, {\sf{}kyconv}, which can be applied to any existing
{\sc{}xspec} model of intrinsic X-ray emission (naturally,
a meaningful combination of the models is the responsibility of
the user). We remind the reader that {\sf{}kyconv} is substantially more
powerful than the usual convolution models in {\sc{}xspec}, which are 
defined in terms of one-dimensional integration over energy bins.
Despite the fact that {\sf{}kyconv}
still uses the standard {\sc{}xspec} syntax in evaluating the observed 
spectrum (e.g.\ {\sf{}kyconv(powerlaw)}), our code
performs a more complex operation. It still performs 
ray-tracing across the disk surface, so the intrinsic model 
contributions are integrated from different radii. The price that 
one has to pay for the enhanced functionality is a higher
demand for computational power.

\begin{figure*}[tbh]
\epsscale{0.8}
\plotone{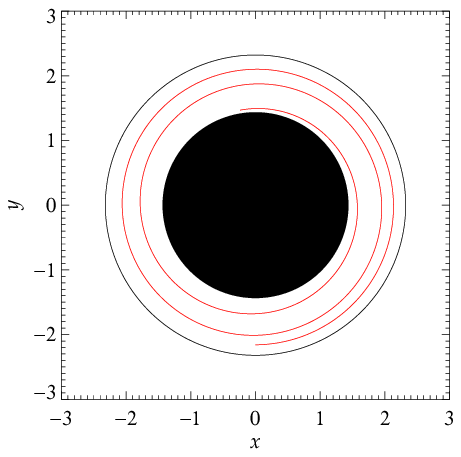}
\epsscale{1.2}
\plotone{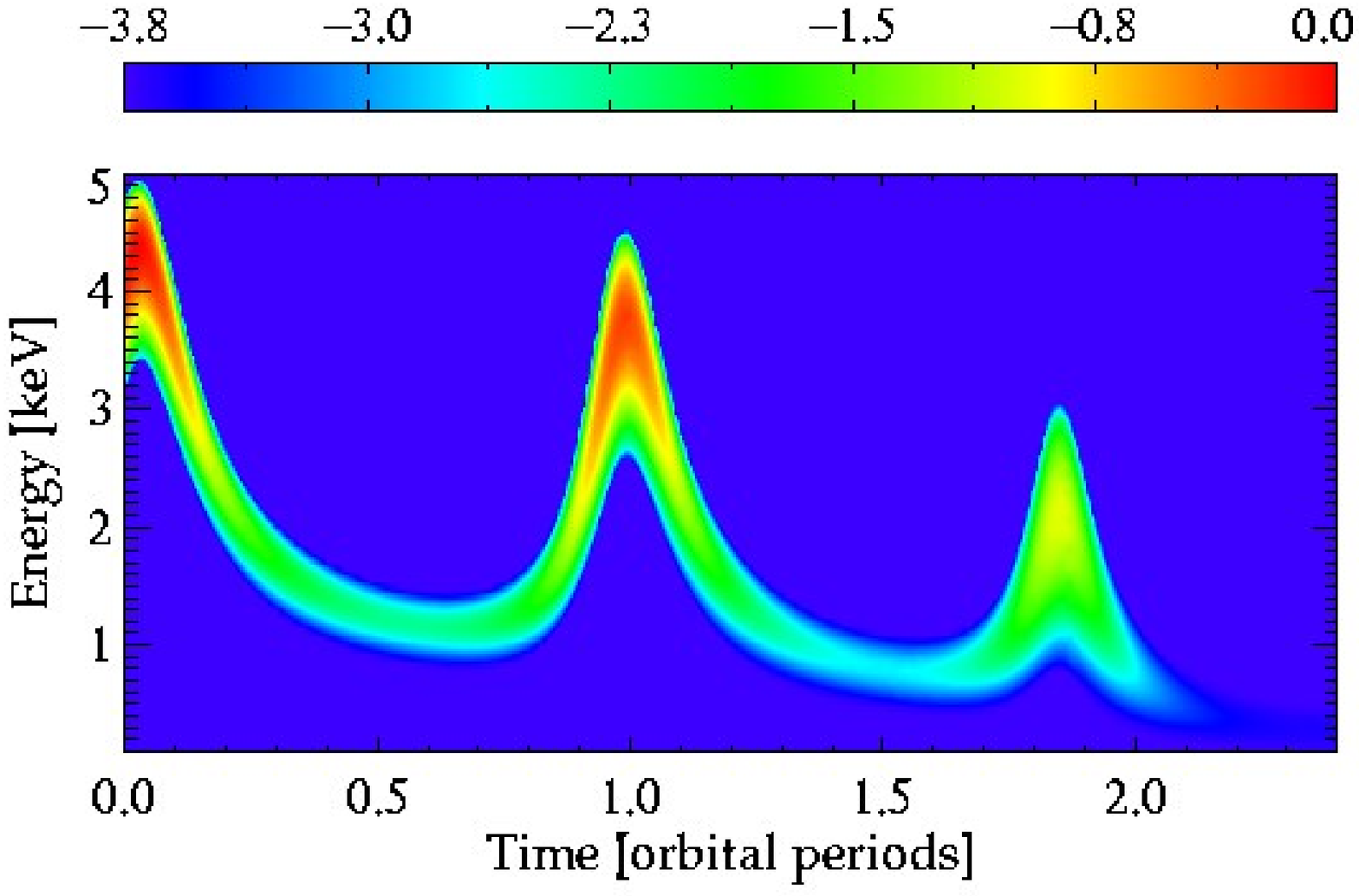}
\caption{Dynamical profile of a line produced by
an in-spiralling spot in free fall. Left: Trajectory
(in the equatorial plane) of the spot,
defined by a constant angular momentum during the infall.
Right: Spectrum of the spot. Energy is on the ordinate, time on
the abscissa. The horizontal range spans the interval of $2.4$ 
orbital periods at the corresponding initial radius, 
$r=0.93\,r_{\rm{}ms}(a)$. Here, the dimension-less angular 
momentum parameter 
of the black hole is $a=0.9$, observer inclination is 
$\theta_{\rm{}o}=45^{\circ}$. The observed photon flux 
is colour-coded (logarithmic scale 
with arbitrary units). Gradual decay of the signal and an 
increasing centroid redshift can be observed as the spot 
completes over two full revolutions and eventually plunges 
into the black hole.}
\label{fig:example2}
\end{figure*}

\begin{figure*}[tbh]
\epsscale{2.2}
\plottwo{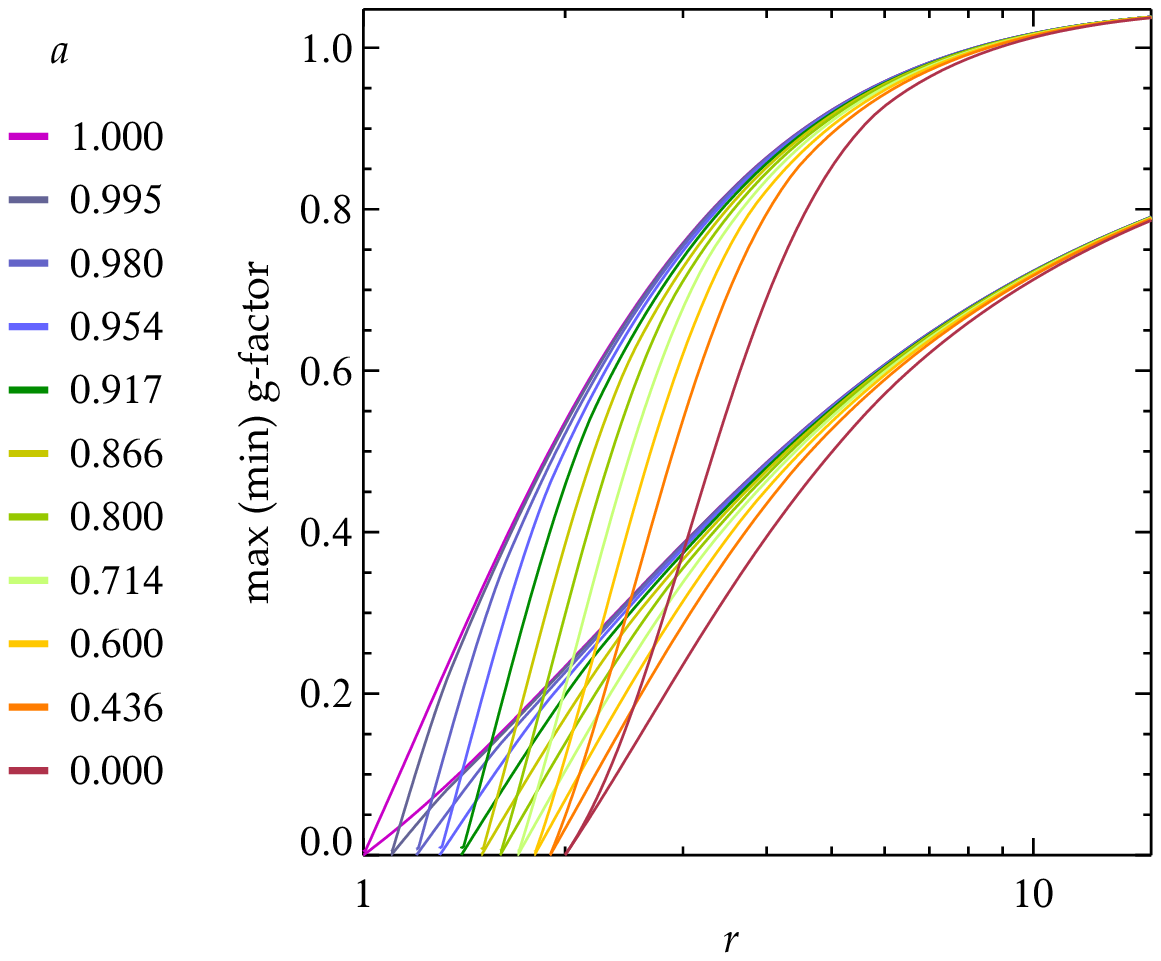}{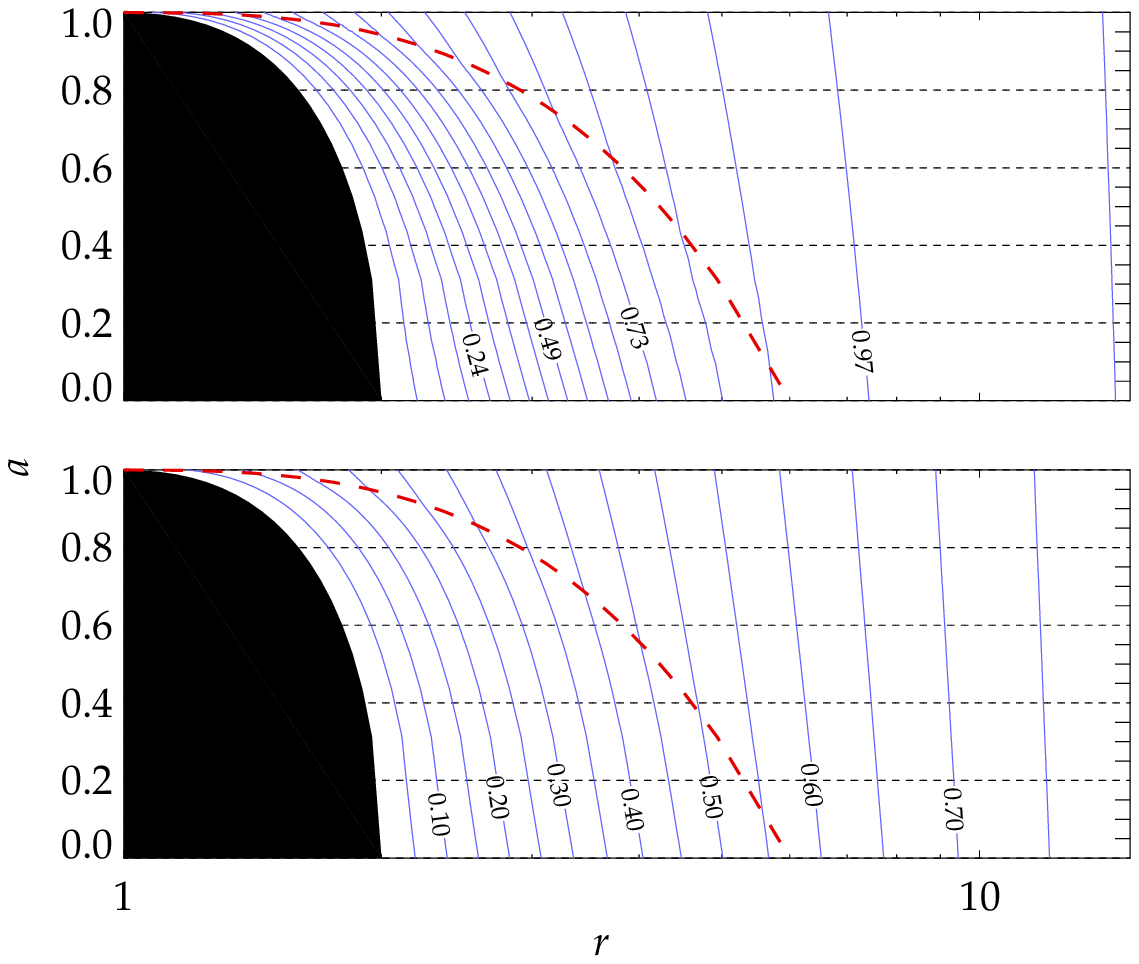}
\caption{Energy shift of photons originating from different
radii in the disk. The dependence on the black-hole
angular momentum, $a$, is shown.
Left: extremal values of the redshift factor $g_{\pm}(a;r)$,
(i.e.\ the maximum and minimum values, corresponding to two bundles of 
neighbouring curves). Each curve corresponds to a fixed value of
$a$ in the range $\langle0,1\rangle$,
encoded by line colours. Right: Corresponding contour lines 
of extremal energy shift factors in the plane of the rotation parameter, $a$,
versus radius $r$ ($g_{+}$ is shown on the top, 
$g_{-}$ in the bottom). Radius is in units of $GM/c^2$. 
The horizon is shown in black. The 
curve $a(r)_{|r=r_{\rm{}ms}}$ (dashed) is also plotted across 
the contour lines. Notice that, in the traditional 
disk-line scheme, no radiation is supposed to originate 
from radii $r<r_{\rm{}ms}$. If this is the case, then
one must assume that all photons originate outside the dashed curve,
and so the effect of frame-dragging is further reduced.
\label{fig:g_minmax}}
\end{figure*}

As mentioned above, time-dependent spectra are expected to
eliminate a great deal of uncertainty in studying 
accretion disks (e.g.\ Done \& Gierli\'{n}ski 2003). 
As an example of using {\sf{}ky} outside of {\sc{}xspec}, 
Fig.~\ref{fig:example2} represents an evolving spectral line 
in a dynamical diagram: a localized spot, free-falling 
from the marginally stable orbit down towards 
the black-hole horizon. Intrinsic emissivity is assumed to fade
gradually with the distance from the spot centre, with
characteristic diameter of the spot $\sim0.4GM/c^2$ 
($\beta=100$ in {\sf{}kyspot}), but it is kept constant
in time. Therefore the figure reflects the effect of 
changing energy shift along the spot trajectory.
The in-spiral motion starts just below the
marginally stable orbit and it proceeds down
to the horizon, maintaining the specific energy and 
angular momentum of its initial circular orbit. 
One can easily recognise that variations in redshift could
hardly be recovered from data if only time-averaged spectrum were
available. In particular, the motion above $r_{\rm{}ms}(a)$ near a 
rapidly spinning black hole is difficult to distinguish from a 
descent below $r_{\rm{}ms}(0)$ in the non-rotating case. 

Before time-resolved spectra are available with sufficient quality,
preliminary considerations can be based on the extremal values, and the
range, of energy shift, $g_{+}$ and $g_{-}$, as well as the relative
time delay which  photons experience when arriving at the observer's
location from different parts of the disk. This is particularly relevant
for some narrow lines whose redshift can be determined more accurately
than if the line is broad (e.g.\ Turner et al.\ 2002; Guainazzi 2003;
Yaqoob et al.\ 2003). Careful discussion of $g_{\pm}$ can, in principle,
circumvent the uncertainties which are introduced by uncertain form of
the intrinsic emissivity of the disk and yet still constrain some of
parameters. Advantages of this technique were pointed out already by
Cunningham (1975) and it was further developed by Pariev, Bromley \&
Miller (2001).

Fig.~\ref{fig:g_minmax} shows the extremal values of the 
redshift factor for $\theta_{\rm{}o}=30^{\circ}$ (which often
seems to be preferred by the observational data). 
These were computed along $r=\mbox{constant}$ 
circles in the disk plane, together with the corresponding contour lines 
of $g_{\pm}=\mbox{constant}$ in the plane $a$ versus $r$. 
In other words, radiation is supposed to originate from radius 
$r$ in the disk, but it experiences a different redshift depending
on the polar angle. Contours of the redshift factor provide a very useful 
and straightforward technique to determine the position of a flare or a 
spot, provided that a narrow spectral line is produced
and measured with sufficient accuracy. The most direct
use of extremal $g$-values would be, if one were able to
measure variations of the line profile from its lowest-energy 
excursion (for $g_{-}$) and highest-energy excursion 
(for $g_{+}$), i.e.\ over a complete cycle. 
Even partial information can help to constrain models (for example,
the count rate is expected to dominate the observed line at the
time that $g=g_{+}$, and so the high-energy peak is 
easier to detect). While
it is very difficult to achieve sufficient precision on the highly
shifted and damped red wing of a broad line, 
prospects for using narrow lines is indeed interesting,
provided that they originate close enough to the black hole and that
sufficient resolution is achieved both in the energy and time domains.

Fig.~\ref{fig:g_minmax} also makes it clear that it is possible in
principle (but intricate in practice) to deduce the $a$-parameter value 
from spectra. It can be seen from the redshift factor that 
the dependence on $a$ is rather small and it quickly becomes
negligible if the light from the source is dominated by contributions 
from $r\gtrsim10$. For example, setting $r=7$ and
$\theta_{\rm{}o}=30^{\circ}$, one finds that the relative difference
between $g_{+}(a=1)$ and $g_{+}(a=0)$ amounts to only
$\sim2\%$. A similar value comes out for
$g_{-}$, and this also roughly corresponds to
the precision which is needed in order to be able to reveal 
the effect of black hole rotation. Notice, however, that this 
is in principle achievable by the {\it{}Chandra} gratings and the 
{\it{}Astro--E2} calorimeter. It may also be achievable by
{\it{}XMM--Newton} with high signal-to-noise data, and of course the
effect would be more pronounced in case of a source seen edge-on.

Other user-defined emissivities can be easily adopted. This can be
achieved either by using the convolution component {\sf{}kyconv} or by
adding a new user-defined model to {\sf{}xspec}. The latter method is
more flexible and faster, and hence recommended. In both approaches, the
ray-tracing routine is linked and used for relativistic blurring.
Naturally, by adding new emissivity laws one invariably introduces
additional freedom (for example, the height $h$ of the primary source in
the lamp-post model). It is worth repeating that caution is always
necessary because the predictive power of the model rapidly decays with
the number of parameters. 

We conclude this section by summarizing the way of using the new model:
Time-dependent spectra are computed by specifying the photon numbers
emitted from the disk, $N_{\rm{}loc}(E,r,\varphi,\mu_{\rm{}loc},t)$, 
where $r$ and $\varphi$ are polar coordinates in the  disk plane,
$\mu_{\rm{}loc}$ is  cosine of the emission angle with respect to the
disk normal, $t$ is Boyer-Lindquist time in the Kerr metric, and $N$ 
comes out in
$\mbox{photons}\,\mbox{cm}^{-2}\,\mbox{s}^{-1}\,\mbox{sr}^{-1}\,\mbox{keV}^{-1}$.
Naturally, mean spectra represent a special case in which time
dependence of the source emissivity is not considered.

\section{Application to MCG--6-30-15}          
The Seyfert~1 galaxy MCG--6-30-15 is a unique source in which the evidence 
of a broad and skewed \fekalfa line has led to a wide acceptance of 
models with an accreting black hole in the nucleus 
(Tanaka et al.\ 1995; Iwasawa et al.\ 1996; Nandra et al.\ 1997; 
Guainazzi et al.\ 1999; Fabian \& Vaughan 2003). Being a nearby
AGN (the galaxy redshift is $z=0.0078$), this source offers 
an unprecedented opportunity to explore directly the pattern of 
the accretion flow onto the central hole.
The Fe K line shape and photon redshifts indicate that a large 
fraction of the emission originates from $r\lesssim10$ ($GM/c^2$).
The mean line profile derived from {\it{}XMM--Newton} observations
is similar to the one observed previously using {\it ASCA}.
The X-ray continuum shape in the hard spectral band was well 
determined from {\it BeppoSAX} data (e.g.\ Guainazzi et al.\ 1999).

\begin{table*}[tbh]
\caption{Spectral fitting results for MCG--6-30-15 using the {\sf{}ky} model.}
\begin{center}
\tabletypesize{\footnotesize}
\begin{scriptsize}
\newlength{\mylen}
\settowidth{\mylen}{{\mbox{$2.1\pm0.2^{\,\dag}$}~}}
\begin{tabular}{c@{}c@{~}c@{~~}c@{~~}c@{}c@{~}c@{~}c@{~}c@{~}c@{~}c@{~}c@{}c@{}c}
\tableline
\multicolumn{1}{c}{{\normalsize{}\#}\rule[-1.5ex]{0mm}{4.5ex}} & 
\multicolumn{1}{c}{{\normalsize$a$}} & 
\multicolumn{1}{c}{{\normalsize$\theta_{\rm{}o}$}} & 
\multicolumn{2}{c}{{\normalsize{}Continuum}} &\multicolumn{1}{c}{}& 
\multicolumn{6}{c}{{\normalsize{}Broad \fekalfa line}} &\multicolumn{1}{c}{}&
\multicolumn{1}{c}{{$\!$\normalsize$\chi^2$}} \\
\cline{4-5}\cline{7-12}\cline{14-14}
   &                        &          & $\Gamma_{\rm{}c}$ & $\alpha_{\rm{}c}$ && $r_{\rm{}in}\!-r_{\rm{}h}$ & $r_{\rm{}b}\!-r_{\rm{}h}$  & $r_{\rm{}out}\!-r_{\rm{}h}$ & $\alpha_{\rm{}in}$ & $\alpha_{\rm{}out}$   & EW                && ~({\sf{}dof})\rule[-1ex]{0mm}{4ex}\\
\tableline
1  & $0.35^{+0.57}_{-0.30}$ & $31.8\pm0.3$ & $2.01\pm0.02$ & $1.0^{+9}_{-1}$   && $5.1\pm0.2$                &      --                    &  $11.4\pm0.8$               &        --            & $3.9\pm0.6$         & $258^{+26}_{-13}$ && {\small$\frac{368.8}{(329)}$}\rule[0ex]{0mm}{4ex}\\
2a & $0.99\pm0.01$          & $40.4\pm0.6$ & $2.03\pm0.02$ & $5.5^{+6}_{-2}$   && $0.67\pm0.04$              & $3.35\pm0.05$              &  $40^{+960}_{-33}$          &  $6.9^{+0.5}_{-0.4}$ & $9.7^{+0.3}_{-0.8}$ & $268\pm13$        && {\small$\frac{308.6}{(330)}$}\rule[0ex]{0mm}{4ex}\\
2b & $0.72^{+0.12}_{-0.30}$ & $28.5\pm0.5$ & $2.01\pm0.01$ & $0.1^{+2}_{-0.1}$ && $0.65\pm0.35$ & \parbox{\mylen}{\mbox{$2.1\pm0.2^{\,\dag}$} \mbox{$7.2\pm0.2$}} &  $48^{+200}_{-25}$           & $8.1^{+1.4}_{-0.9}$ & $4.9^{+0.4}_{-0.3}$ & $241^{+13}_{-10}$ && {\small$\frac{313.5}{(330)}$}\rule[0ex]{0mm}{4ex}\\
2c & $0.25\pm0.03$          & $27.6\pm0.6$ & $1.97\pm0.02$ & $3.1^{+0.3}_{-0.1}$ && $1.23\pm0.06$            & $4\pm0.02$                 &  $109^{+20}_{-10}$          & $9.2\pm0.2$          & $3.1\pm0.1$         & $267\pm10$        && {\small$\frac{313.9}{(330)}$}\rule[0ex]{0mm}{4ex}\rule[-3ex]{0mm}{7ex}\\
\tableline
\end{tabular}
\end{scriptsize}

\vspace*{2mm}\par{}
{\parbox{0.95\textwidth}{\footnotesize{}
Best-fitting values of the important parameters and their 
statistical errors for models \#1--2, described in the text.
The models include broad \fekalfa and \fekbeta emission
lines, a narrow Gaussian line at $\sim 6.9$~keV, and 
a Compton-reflection continuum from a relativistic disk. 
These models illustrate different assumptions 
about intrinsic emissivity of the disk (the radial
emissivity law need not to be a simple power law, but axial
symmetry has been still imposed here).
The inclination angle $\theta_{\rm{}o}$ 
is in degrees, relative to the rotation axis; radii are expressed 
as an offset from the horizon (in $GM/c^2$);
the equivalent width, EW, is in electron volts.\\
$^{\dag}$~Two values of the transition radius define the interval 
$\langle{}r_{\rm{}b-},r_{\rm{}b+}\rangle$ where reflection is 
diminished.}}
\label{tab:best-fit-models}
\end{center}
\end{table*}

\begin{figure*}[tbh]
\epsscale{2.2}
\plottwo{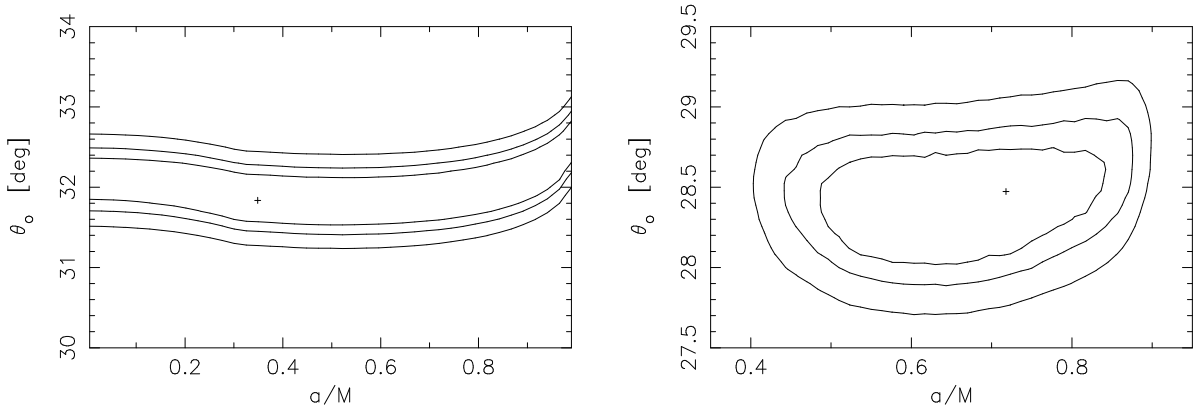}{f3b.eps}
\caption{Confidence contours around the best-fitting parameter 
values (indicated by a cross). Left: the case of a single 
line-emitting region (model~\#1 with zero emissivity for 
$r<r_{\rm{}ms}$). Right:
the case of a non-monotonic radial emissivity, model~\#2b.
The joint two-parameter contour levels
for $a$ versus $\theta_{\rm{}o}$ correspond to 
68\%, 95\% and 99\% confidence.}
\label{fig:confcont1}
\end{figure*}

To illustrate the new {\sf{}ky} model capabilities,
we used our code to analyze the mean EPIC PN spectrum which we 
compiled from the long {\it{}XMM--Newton} 2001
campaign (e.g., as described in Fabian et al.\ 2002).
The data were cleaned and reduced using standard data reduction 
routines, employing 
SAS version 5.4.1.\footnote{See {\sf{}\url{http://xmm.vilspa.esa.es/sas/}}.}
We summed the EPIC PN data from five observations
made in the interval 11 July 2001 to 04 August 2001,
obtaining a total good exposure time of $\sim290$~ks (see
Fabian et al.\ 2002 for details of the observations).
The energy range was restricted to $3$--$10$~keV with $339$ energy
bins, unless otherwise stated, and models were fitted by minimizing 
the $\chi^{2}$ statistic. Statistical errors quoted correspond to 
$90$\% confidence for one interesting parameter 
(i.e.\ $\Delta\chi^{2}=2.706$), unless otherwise stated.

No absorber was taken into account; the assumption
here is that any curvature in the spectrum above 3~keV is
not due to absorption, and only due to the Fe K line. 
We remind the reader that the Fe~K emission line dominates
around the energy $6$--$7$~keV, but it has been supposed to stretch
down to $\sim3$~keV or even further.
We emphasize that our aim here is to test the hypothesis that 
{\it{}if all of the curvature is entirely due to the broad
Fe~K feature, is it possible to constrain $a$
of the black hole?} Obviously, if the answer to this is `no', it
will also be negative if some of the spectral curvature
between $\sim3$--$5$~keV is due to processes other than
the Fe~K line emission. 

We considered \fekalfa and \fekbeta
iron lines (with their rest-frame energy fixed at $E_{\rm{}loc}=6.400$ 
and $7.056$~keV, respectively), a narrow-line feature 
(observed at $E_{\rm{}obs}\sim6.9$~keV) modeled as a Gaussian with center
energy, width, and intensity free, plus a power-law
continuum. We used a superposition of two {\sf{}kygline} models 
to account for the broad relativistic K$\alpha$ and K$\beta$ lines, 
{\sf{}zgauss} for 
the narrow, high-energy line (which we find to be centered at
$6.86$~keV in the source frame, slightly redshifted relative
to $6.966$~keV, the rest-energy of Fe~Ly$\alpha$), 
and {\sf{}kyhrefl} for the Comptonized continuum convolved 
with the relativistic kernel. We assumed that the high-energy narrow
line is non-relativistic. Note that an alternative interpretation of the
data (as pointed out by Fabian et al.\ 2002) is that there
is not an emission line at $\sim 6.9$~keV, but He-like resonance 
absorption at $\sim 6.7$~keV. This would not affect our conclusions.
The ratio of the intensities of the two relativistic lines was fixed at 
$I_{\rm{K\beta}}/I_{\rm{K\alpha}}=0.1133$ ($=17/150$), and the iron 
abundance for the Compton-reflection continuum was assumed to be three times
the solar value (Fabian et al.\ 2002). We used the angles 
$\theta_{\rm{}min}=0^{\circ}$ and $\theta_{\rm{}max}=90^{\circ}$ for 
local illumination in {\sf{}kyhrefl} (these values are equivalent to central 
illumination of an infinite disk).
The normalization of the Compton-reflection continuum relative
to the direct continuum is controlled by the
effective reflection `covering factor', $R_{\rm{}c}$, which is the
ratio of the actual reflection normalization to that expected
from the illumination of an infinite disk (see Appendix).
Also $r_{\rm{}out}$ (outer radius of the disk), $\alpha_{\rm{}c}$ 
(slope of power-law continuum radial emissivity in the {\sf{}kyhrefl}
model component), as well as $R_{\rm{}c}$, were included among the
free parameters, but we found them to be only very poorly constrained.
Weak constraints on $\alpha_{\rm{}c}$ and 
$R_{\rm{}c}$ are actually expected in this model for two reasons.
Firstly, in these data the continuum is indeed rather featureless. Secondly,
the model continuum was blurred with the relativistic kernel of {\sf{}kyhrefl}, 
and so the dependence of the final spectrum on the exact form of
emissivity distribution over the disk must be quite weak.

\textit{Model \#1.} 
Using a model with a plain power-law radial emissivity
on the disk, we obtained the best-fitting values for the 
following set of parameters: $a$ of the black hole, disk 
inclination angle $\theta_{\rm{}o}$, radii of the line-emitting region,
$r_{\rm{}in}<r<r_{\rm{}out}$, the corresponding radial emissivity 
power-law index $\alpha$ of the line emission, as well as the radial 
extent of the continuum-emitting region, its photon index $\Gamma_{\rm{}c}$, 
and the corresponding $\alpha_{\rm{}c}$ for the continuum.
Notice that $\Gamma_{\rm{}c}$ and $\alpha_{\rm{}c}$ refer 
to the continuum component {\it{}before\/} the relativistic 
kernel was applied to deduce the observed spectrum.
As a result of the integration across the disk, the model 
weakly constrains these parameters. 

There are two ways to interpret this. A model which is
over-parameterized is undesirable from the point of view of deriving 
unique model parameters from modeling the data. However, another
interpretation is that a model with a greater number of parameters may
more faithfully  reflect the real physics and it is the actual physical
situation which leads to degeneracy in the model parameters. The latter
implies that some model parameters can never be constrained uniquely,
regardless of the quality of the data. In practice one must apply both
interpretations and assess the approach case by case, taking into
account the quality of the data, and which parameters can be constrained
by the data and which cannot. If preliminary fitting shows that large
changes in a parameter do not affect the fit, then that parameter can be
fixed at some value obtained by invoking physically reasonable arguments
pertaining to the situation. 

We performed various fits with the inner edge tied
to the marginally stable orbit and also fits where $r_{\rm{}in}$
was allowed to vary independently. Free-fall motion with constant
angular momentum was assumed below $r_{\rm{}ms}$, if the emitting region
extended that far. Table~\ref{tab:best-fit-models}
gives best-fitting values of the
key relativistic line and continuum model parameters
for the case in which $r_{\rm{}in}$ and $a$ were independent.

Next, we froze some of the parameters at their best-fit 
values and examined the $\chi^2$ space by varying the remaining 
free parameters. That way we constructed joint confidence contours
in the plane $a$ versus $\theta_{\rm{}o}$
(see left panel of Fig.~\ref{fig:confcont1}).
These representative plots demonstrate that $\theta_{\rm{}o}$ 
appears to be tightly constrained, while $a$ is allowed to vary 
over a large interval around the best fit, extending down to $a=0$. 

\textit{Model \#2.} 
We explored the possibility that the broad-line 
emission does not conform to a unique power-law radial
emissivity but that, instead,
the line is produced in two concentric rings (a `dual-ring' model). 
This case can be considered as a toy model for a more complex
(non-power law) radial dependence of the line emission than the 
standard monotonic decline, which we represent here by
allowing for different values of $\alpha$ in the inner and outer
rings: $\alpha_{\rm{}in}$ and $\alpha_{\rm{}out}$.
Effectively, large values of the power law index 
represent two separate rings.
The two regions are matched at the transition radius, $r=r_{\rm{}b}$, and
so this is essentially a broken power law. We explored both the case of 
continuous and discontinuous line emissivity at $r=r_{\rm{}b}$. Notice that 
the double power-law emissivity arises naturally in the lamp-post model
(Martocchia et al.\ 2000) in which the disk irradiation and 
the resulting \fekalfa reflection are substantially anisotropic
due to fast orbital motion in the inner ring. Although the lamp-post
model is very simplified in several respects, namely the way in which the
primary source is set up on the rotation axis, one can expect fairly
similar irradiation to arise from more sophisticated schemes of
coronal flares distributed above the disk plane. 
Also, in order to provide a physical picture of the steep emissivity
found in {\it{}XMM--Newton} data of MCG--6-30-15,
Wilms et al.\ (2001) invoked strong magnetic stresses acting 
in the innermost part of the system, assuming that they are able to 
dissipate a considerable amount of energy in the disk     
at very small radii. Intense self-irradiation 
of the inner disk may further contribute to the effect. 

This more complex model is consistent
with the findings of Fabian et al.\ (2002). Indeed, the
fit is improved relative to 
models with a simple emissivity law because the 
enormous red wing and relatively sharp core
of the line are better reproduced thanks to 
the contribution from a highly redshifted inner disk
(see Table~\ref{tab:best-fit-models}, model~\#2a).
For the same reason that the more
complex model reproduces the line core along
with the red wing well, the model prefers higher values of
$a$ and $\theta_{\rm{}o}$ than what we found for 
the case \#1. Notice that $a\rightarrow1$
implies that all radiation is produced above $r_{\rm{}ms}$.
Maximum rotation is favoured with
both $a$ and $\theta_{\rm{}o}$ 
appearing to be tightly constrained near their 
best-fit values. Likewise for the continuum radial emissivity
indices. There is a certain freedom
in the parameter values that can be accomodated by this model.
By scanning the remaining parameters, 
we checked that the reduced $\chi^2\sim1$ can be achieved
also for $a$ going down to $\sim0.9$ and 
$\theta_{\rm{}o}\sim37^{\circ}$. This conclusion is also consistent with
the case for large $a$ in Dabrowski et al.\ (1997); however we actually
do not support the claim that the current data {\it{}require}
a large value of $a$. As shown below, reasonable
assumptions about intrinsic emissivity can fit the data with small $a$
equally well.

In model \#2a, small residuals remain near $E\sim4.8$~keV
(at about the $\sim1$\% level), 
the origin of which cannot easily be clarified with 
the time-averaged data that we employ now. The
excess is reminiscent of a Doppler horn typical of relativistic line 
emission from a disk, so it may
also be due to \fekalfa emission which is locally enhanced
on some part of the disk. We were able to reproduce the peak 
by modifying the emissivity at the transition radius,
where the broken power-law emissivity changes its slope (model~\#2b).
We can even allow non-zero emissivity below $r_{\rm{}ms}$ (the inner ring) 
with a gap of zero emissivity between the outer edge of the inner 
ring and the inner edge of the outer one. The inner ring, 
$r_{\rm{}in}{\leq}r{\leq}r_{\rm{}b-}$, contributes to the red tail
of the line while the outer ring, $r_{\rm{}b+}{\leq}r{\leq}r_{\rm{}out}$, 
forms the main body of the broad line. The resulting plot of
joint confidence contours of $\theta_{\rm{}o}$ versus $a$ is shown 
in Fig.~\ref{fig:confcont1} (right panel)
In order to construct the confidence contours we scanned a broad 
interval of parameters, $0<a<1$ and $0<\theta_{\rm{}o}<45^{\circ}$; 
here, detail is shown only around the minimum $\chi^2$ region.

\begin{figure*}[tbh]
\epsscale{2.2}
\plottwo{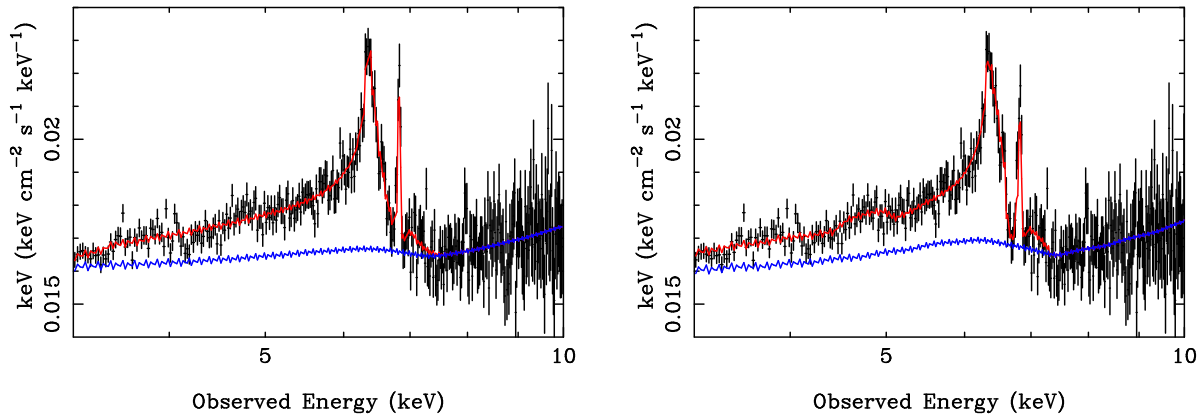}{f4b.eps}
\caption{Spectrum and best-fitting model for the {\it XMM--Newton} 
data for MCG--6-30-15 in which the Fe~K line originates in a dual-ring. 
The models 2a (left) and 2b (right) are shown in comparison.
See Table~\ref{tab:best-fit-models} for the model parameters.
The continuum component ({\sf{}kyhrefl}) is also plotted.
The data points are not unfolded: the spectrum in these units was
made by multiplying the ratio of measured counts to the counts
predicted by the best-fitting model and then this ratio was multiplied by
the best-fitting model and then by $E^{2}$.}
\label{fig:dualline3-10kev_v3}
\end{figure*}

Two examples of the spectral profiles are shown in
Fig.~\ref{fig:dualline3-10kev_v3}. It can be seen that the overall
shapes are very similar and the changes concern mainly the red wing of
the profile. For completeness we also fitted several modifications of
the model \#2 and found that with this type of disk emissivity (i.e.\
one which does not decrease monotonically with radius) we can still
achieve comparably good fits which have small values of $a$. The case
\#2c in Table~\ref{tab:best-fit-models} gives another example. 
This shows that one cannot draw firm conclusions about $a$ (and some of the 
other model parameters) based on the redshifted part of the line
using current data. The data used here represent the highest signal-to-noise
relativistically broadened Fe K line profile yet available for any AGN.

The differences in $\chi^2$ between model \#1 and models 
\#2a, 2b, and 2c  are very large ($\sim60$) for only one 
additional free parameter, so all variations of model 2 shown
are better fits than the case \#1. We note that formally, model \#2a 
and \#2c appear to constrain $a$ much more tightly than model \#1.
The issue here is that the values of $a$ can be completely different, 
with the statistical errors on $a$ not overlapping (compare, for example,
model \#2a and \#2c in Table~\ref{tab:best-fit-models}). 
This demonstrates that, at least for the parameter $a$, 
it is not simply the statistical error only which determines
whether a small $a$ value is or is not allowed by the data.
We confirmed this conclusion with more complicated models obtained
in {\sf{}ky} by relaxing the assumption of axial symmetry;
additional degrees of freedom do not change the previous results.

One should bear in mind a well-known technical difficulty 
which is frequently encountered while scanning
the parameter space of complex models 
and producing confidence contour plots similar to 
Fig.~\ref{fig:confcont1}. That is, in a rich parameter space
the procedure may be caught in a local minimum which produces
an acceptable statistical measure of the goodness of fit
and appears to tightly constrain parameters near
the best-fitting values. However, manual searching revealed
equally acceptable results in rather remote 
parts of the parameter space. Indeed, as the
results above show, we were able to find 
acceptable fits with the central
black hole rotating either slowly or rapidly (in terms of the $a$
parameter). This fact is not in contradiction with previous 
results (e.g.\ Fabian et al.\ 2002) because we assumed a 
different radial profile of intrinsic emissivity, but it indicates 
intricacy of unambiguous determination of 
model parameters. We therefore need more observational 
constraints on realistic physical mechanisms
to be able to fit complicated models to actual data with
sufficient confidence (Ballantyne, Ross \& Fabian 2001;
Nayakshin \& Kazanas 2002; R\'o\.za\'nska et al.\ 2002).

We have seen that the models described above are able 
to constrain parameters with rather different 
degrees of uncertainty.
It turns out that the more complex type \#2 models (i.e.\ those
that have radial emissivity profiles which are
non-monotonic or even have an appreciable contribution from 
$r<r_{\rm{}ms}$) provide better fits to the
data but a physical interpretation is not obvious.
Ballantyne, Vaughan \& Fabian (2003) also deduce a
dual-reflector model from the same data and propose that the
outer reflection is due to the disk being warped or flared
with increasing radius.

\section{Discussion and Conclusions}
We have presented an extended computational scheme which can be
conveniently employed to examine predicted X-ray spectral features from
black-hole accretion disks. As mentioned above, a one-dimensional
version of the code basically reproduces previous methods, with the
addition of the option to vary more  parameters (e.g.\ $a$). The
current two-dimensional  version also accomodates non-axisymmetric and
time-evolving emissivity from a geometrically thin disk. The new tool
facilitates accurate comparisons  between model spectra and actual data.
It provides also  polarimetric information. One can even substitute
other tables for the Kerr metric tables to explore different specetime.
However, we defer detailed discussion of  these latter capabilities to a
future paper. Furthermore, work is in progress to include realistic
emissivities relevant to different physical situations, in order to
allow for the intrinsically three-dimensional  geometry of the sources
(e.g.\ an optically-thin corona above the disk), and to parallelize the
code.

Recently, several research groups have embarked on projects to compare
theoretical predictions of intrinsic emissivity of accretion disks 
with observational data obtained by X-ray satellites. This approach 
offers a fascinating opportunity to explore processes near 
black-hole sources, both AGNs and BHCs, and to estimate 
physical parameters of black holes themselves. In order to limit
the number of unknown model parameters and to alleviate the computational
burden of the fit procedure, simplifications on the model emissivity
(for example, stationarity and/or axial symmetry)
have been often imposed.\footnote{Traditionally,
Occam's razor is invoked to justify simplifying assumptions. However,
this way of reasoning is not sufficiently substantiated here
in view of the fact that accretion flows are normally found
to be very turbulent and fluctuating rather than smooth and steady.
Put in other words, the assumption of an axisymmetric and stationary
disk may provide a successful fit to the time averaged spectrum,
but it is not satisfactory because we know that more complicated
models {\it{}must} be involved on physical grounds. (``Things should be 
made as simple as possible---but not too simple\ldots'')}
Naturally, fitting data with complex models is computationally more 
demanding.

To summarize, we have illustrated the code
by following common practice and employing 
minimisation of $\chi^2$ to find the best-fitting parameter values
for the highest signal-to-noise Fe K line profile
available to date---from the {\it{}XMM--Newton} campaign of 
the Seyfert~1 galaxy MCG--6-30-15. 
The new model can vary more of the parameters
than previous variants of the 
disk-line scheme allowed, which seems to be necessary in view of the 
fact that simple prescriptions for intrinsic emisivity 
(axisymmetric, radially monotonic, etc.) are not adequate to 
reflect realistic simulations of accretion flows. It is without much 
surprise that a complex model exhibits an intricate $\chi^2$ space in 
which ambiguities arise with multiple islands of acceptable parameters. 
The mean {\it{}XMM--Newton} spectrum of MCG--6-30-15 analyzed here 
justifies rejection of the often-used 
plain power-law emissivity that depends solely on radius 
from the black hole. However, we cannot 
arbitrate the issue of non-monotonic emissivity versus a non-axisymmetric 
one. Due to the inherent degeneracy in time-averaged spectra, both alternatives 
can reproduce these data with reasonable parameters. 
In other words, $a$ of the black hole cannot be unambiguously determined 
if complex (realistic) emissivities are adopted. 

It is clear that the relativistic line profiles
affected by strong gravity are such that it is very difficult to recover
unambiguous information about the key parameters, such as black-hole
spin, disk inclination angle, and inner disk radius. This is because
even with increased throughput, energy resolution and signal-to-noise,
the smooth, featureless tails of the time-averaged line profiles do not
retain sufficient information to break the degeneracies between the
dependences on the black-hole, accretion disk, and line emissivity
parameters. 

Higly specific spectral shapes could actually be provided by
contributions to the line profile from higher-order images, whose form
inherits unique information about space-time of the black hole. It may
appear rather unlikely that higher-order images could be discovered in
spectra of AGNs, because this would require to see an unobscured inner
disk at large (almost edge-on) inclination. 

As the data improve, the important measurable
quantities that will constrain parameters when compared with the models
are: (i)~the slope of the red wing, (ii)~the energy of the cut-off of
the blue side of the line profile, and (iii)~the detailed shape of the
blue peak (see Figs.\ \ref{fig:example0}--\ref{fig:example3} and further 
discussion in the Appendix). The extreme part of the red wing will
not be as useful because it will always be difficult to tell where the
line emission joins the continuum no matter how good the data are, and
also because there is often no prominent sharp feature in the red wing
since the red Doppler peak is too smeared out when the line emission
extends too close to the black hole. In addition, time variability will
be extremely important. No matter how, or if, the line emission responds
to the continuum, we know that the disk inclination and black-hole spin
very likely do not change during different snapshots. Therefore any
differences in the snapshot line profiles must be due only to changes in
the line emissivity, in terms of its radial and azimuthal distribution
and/or its radial extent ($r_{\rm{}in}$ and $r_{\rm{}out}$). Combining these
approaches and collecting enough snapshots may provide sufficient
information to uniquely determine one of the parameters, such as
black-hole spin.

\acknowledgments
We thank A.~Ptak for help with the {\it{}XMM--Newton} data reduction and
use of the {\sf{}Xassist} software package.\footnote{See 
{\sf{}\url{http://xassist.pha.jhu.edu/}}.} We thank A.~C. Fabian for
helpful discussions and for providing us with a relativistic blurring
routine which we used to test our results in the case of maximum
black-hole rotation, B.~Czerny for discussion about the importance of
different darkening laws, and an anonymous referee for useful comments
on the first version of the manuscript.  A.~Martocchia and G.~Matt
provided us with tabular data for the emissivity in the lamp-post model,
and these have been also included in our model. We thank Kim Weaver for
her support, in many ways, of this work. We gratefully acknowledge 
financial support from GACR grant 205/03/0902 and GAUK 2004 (VK), 
GACR 202/02/0735 and GAUK 166/2003 (MD), 
and from NASA grants NCC5-447 and NAG5-10769 (TY).
This research made use of the HEASARC online data archive services,
supported by NASA/GSFC.

\appendix
\onecolumn
\section{The Numerical Code: Tests and Examples}
\label{appa}

\begin{figure*}[!tbh]
\epsscale{0.7}
\plotone{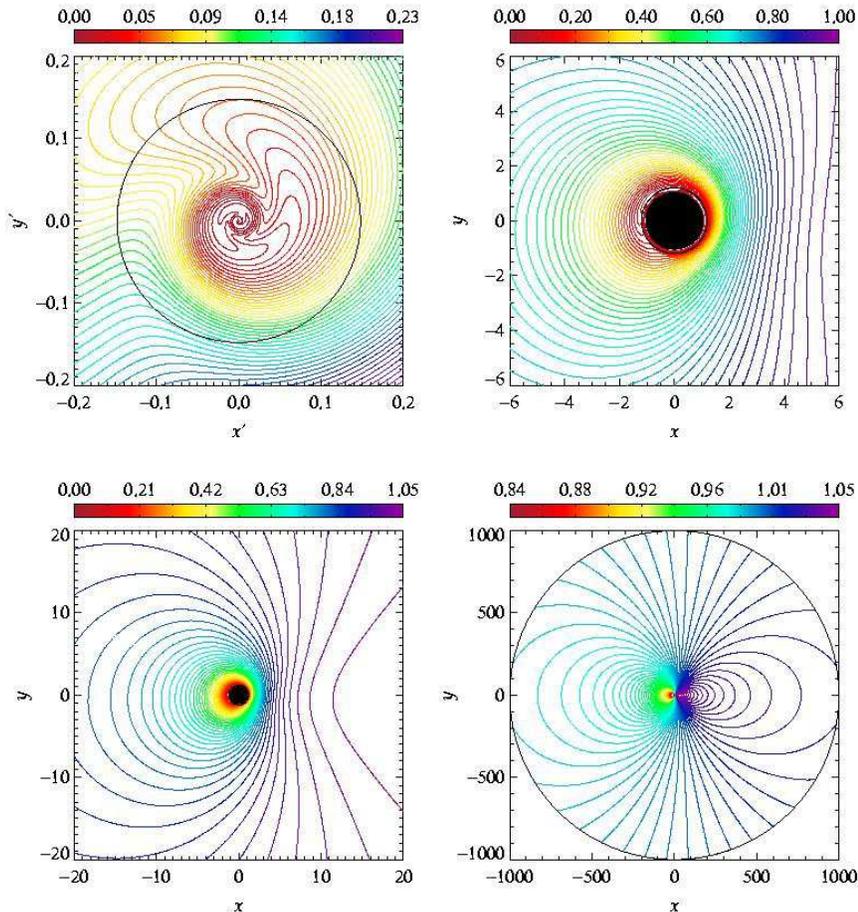}
\caption{Graphical representation of the contents of 
the data tables for the {\sf{}ky} model used in the 
computations. Several sets of contour plots capture the equatorial 
plane with an increasing resolution near the black-hole horizon.
Here we show the effect of energy shift $g{\equiv}g(r,\varphi)$.
Lengths are in $GM/c^2$ 
(radius up to $r=10^3$ in the coarsest grid). 
Values are encoded by a colour scale, as indicated in each
graph. Key parameters are: $\theta_{\rm{}o}=30^{\circ}$, 
and $a=0.9987$ (corresponding to $r_{\rm{}h}\dot{=}1.05$, 
$r_{\rm{}ms}\dot{=}1.2$). Clock-wise distortion of the contours is 
due to frame-dragging near a rapidly rotating Kerr black hole, 
and it is clearly visible in Boyer-Lindquist coordinates here. 
High accuracy of the tables has been achieved by eliminating 
this influence, with the aid of an appropriate transformation. 
In the upper left panel, corresponding to the
region closest to the black hole, 
the horizon has been brought to the origin by
the transformation of coordinates described in the text. 
\label{fig:example6a}}
\end{figure*}

\begin{figure*}[!tbh]
\epsscale{0.7}
\plotone{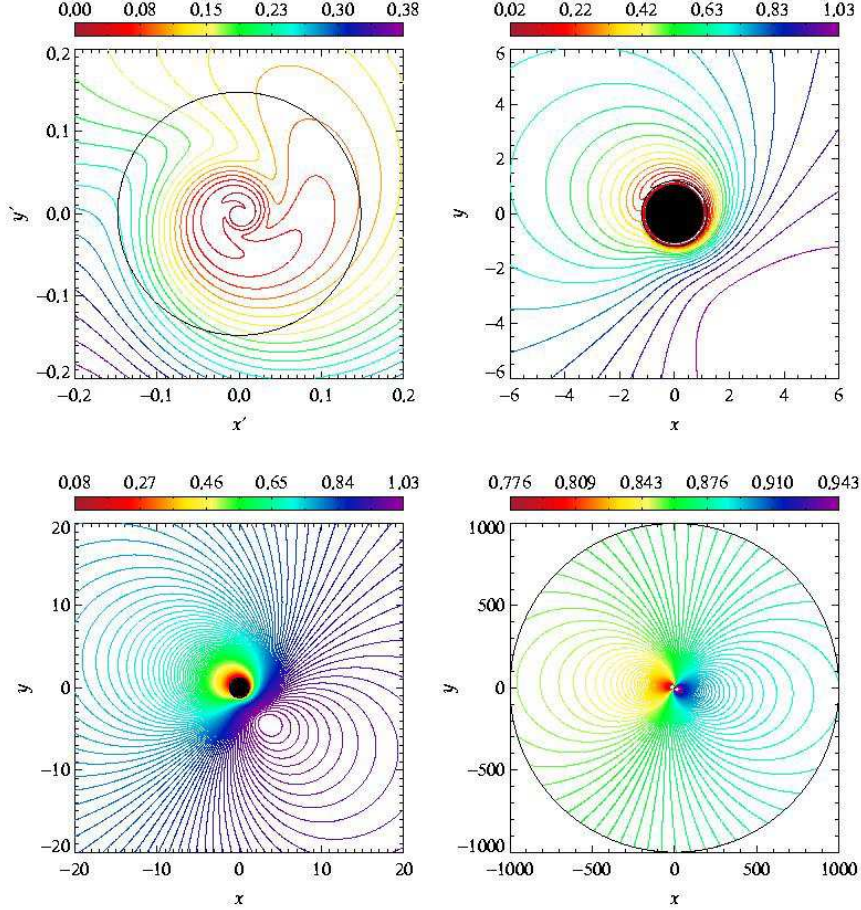}
\caption{The same as in previous figure, but now showing the gravitational 
lensing effect in the disk. Contours correspond to lines of
constant magnification. They were computed by solving geodesic deviation
equations in the Kerr metric.
\label{fig:example6b}}
\end{figure*}

\begin{figure*}[tbh]
\epsscale{1}
\plottwo{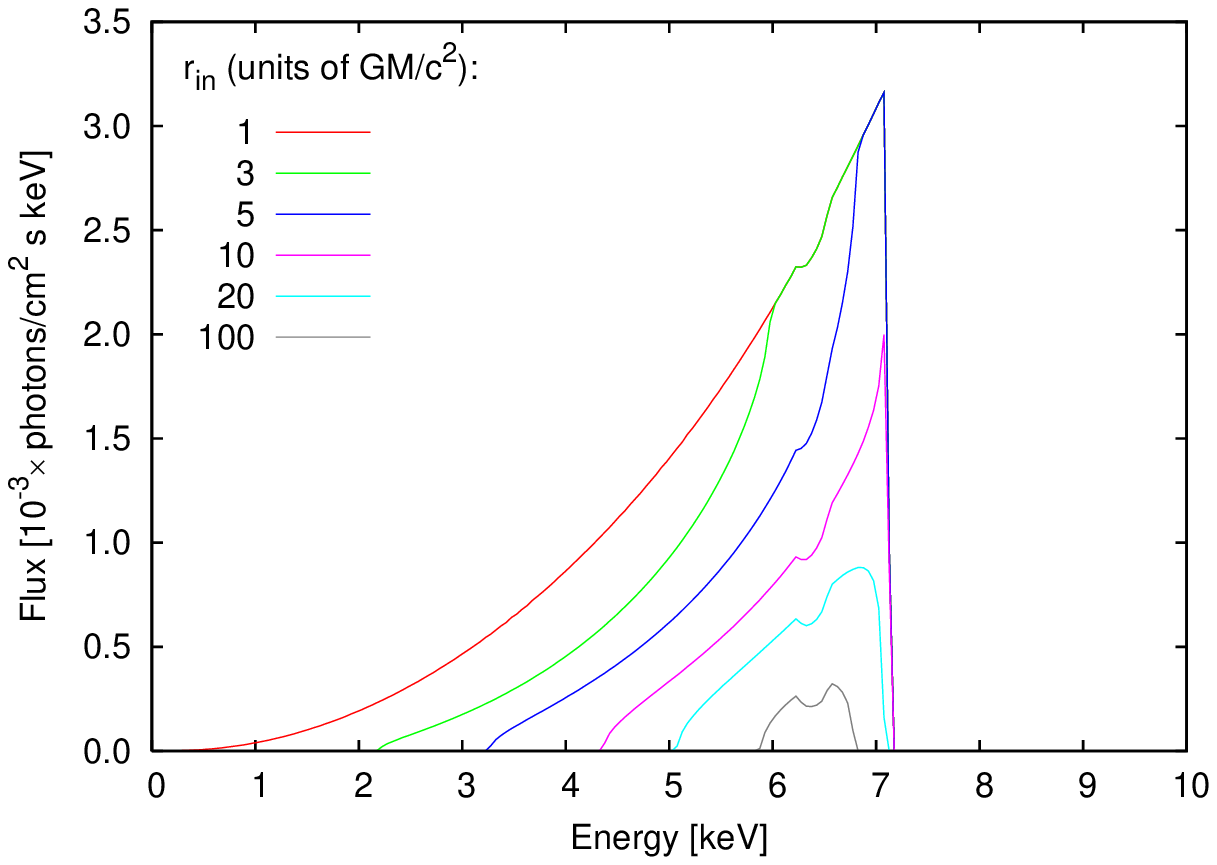}{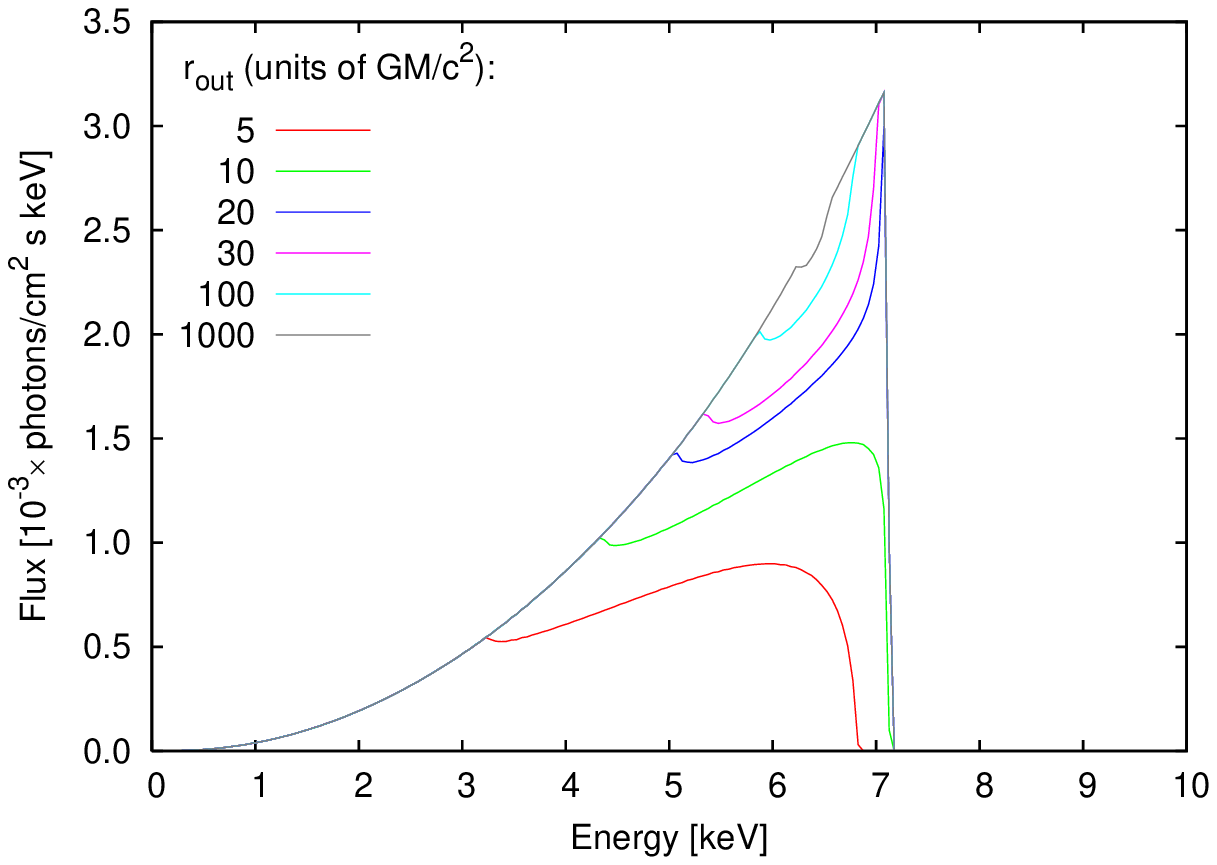}
\caption{Comparative examples of simple line profiles,
showing a theoretical line ($E_{\rm{}loc}=6.4\,$keV) with relativistic
effects originating from a black-hole accretion disk. Different sizes of
the annular region (axially symmetric) have been considered, assuming 
that the intrinsic emissivity obeys a power law in the radial direction 
($\alpha=3$). Resolution of the line-emitting region was
$n_r{\times}n_\varphi=3000\times1500$ with a non-equidistant layout of the
grid in Kerr ingoing coordinates, as described in the text. Left:
dependence on the inner edge. Values of $r_{\rm{}in}$ are indicated in the 
plot (the outer edge has been fixed at the maximum
radius covered by our tables, $r_{\rm{}out}=10^3$). Right: dependence
on $r_{\rm{}out}$ (with the inner edge at horizon,
$r_{\rm{}in}=r_{\rm{}h}$). Other key parameters are: 
$\theta_{\rm{}o}=45^{\circ}$, $a=1.0$. Locally isotropic emission was
assumed in the disk co-rotating frame.}
\label{fig:example0}
\end{figure*}

\begin{figure*}[tbh]
\epsscale{1}
\plottwo{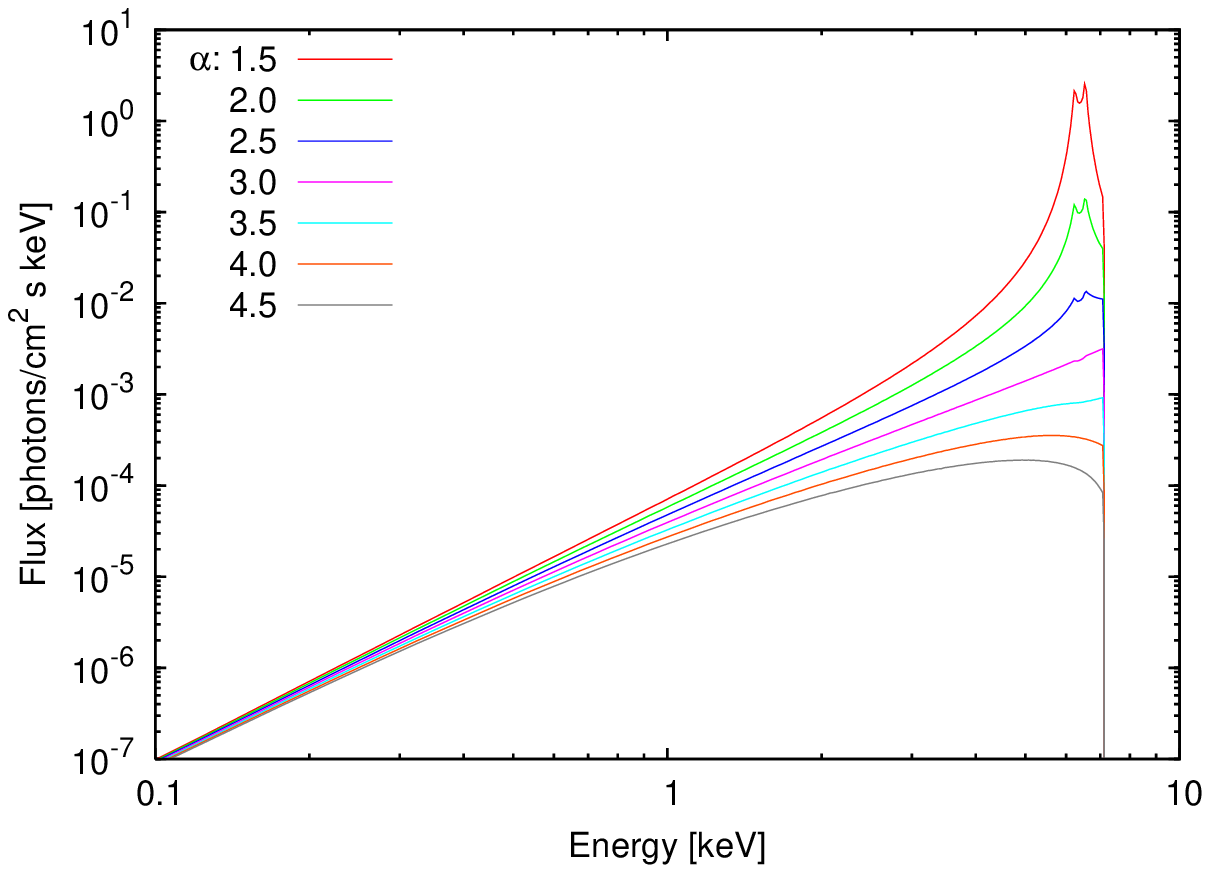}{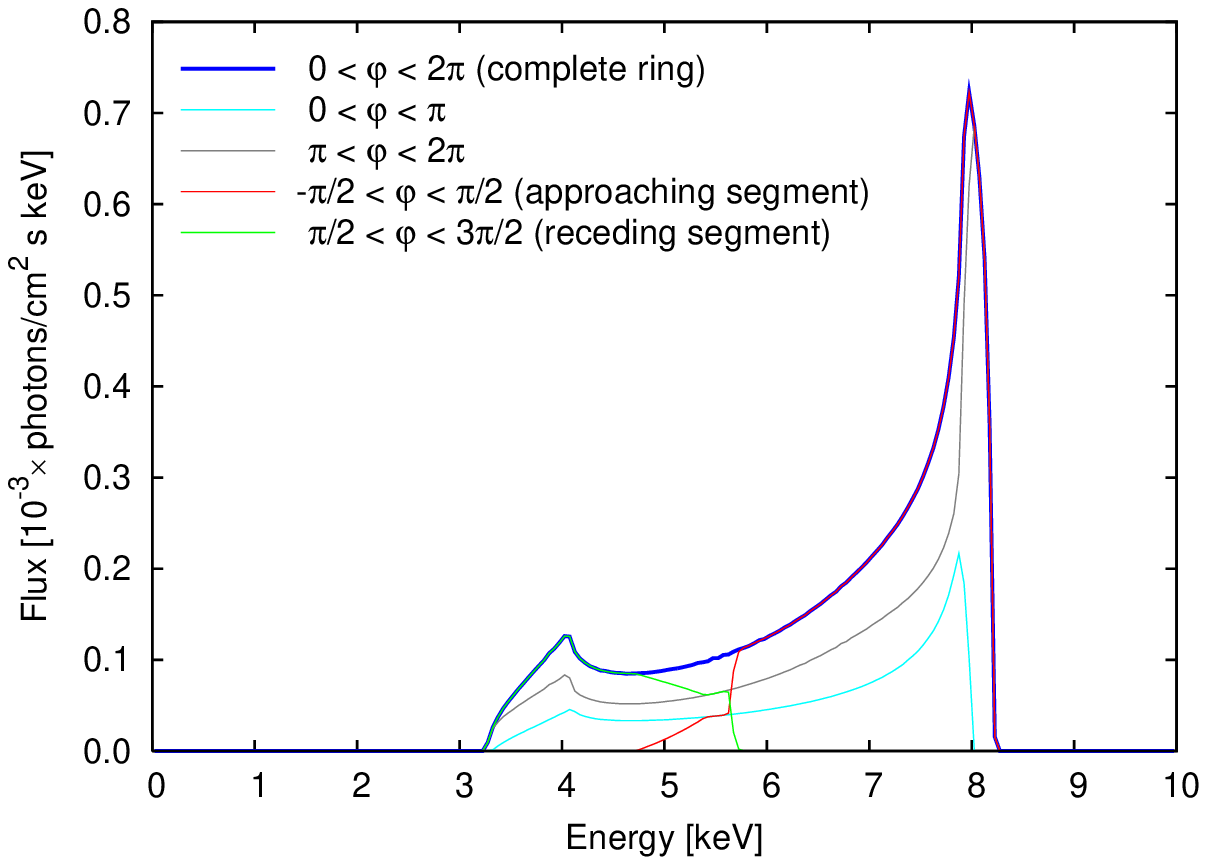}
\caption{More calculated line profiles, as in the previous figure. Left: Line
profiles for different values of $\alpha$. Notice the enhanced red tail
of the line when the intrinsic emission is concentrated to the center of
the disk. Right: Line emission originating from four different azimuthal
segments of the disk. This plot can serve as a toy model of
non-axisymmetric emissivity or to illustrate the expected  effects of
disk obscuration. Obviously, the receding segment of the disk
contributes mainly to the low-energy tail of the line while the
approaching segment constitutes the prominent high-energy peak. These
two plots illustrate a mutual interplay between the effect of  changing
$\alpha$ and the impact of obscuration, which complicates interpretation
of time-averaged spectra. The radial range is $r_{\rm{}h}<r<10^3$ in both
panels.}
\label{fig:example1}
\end{figure*}

\begin{figure*}[tbh]
\epsscale{1}
\plottwo{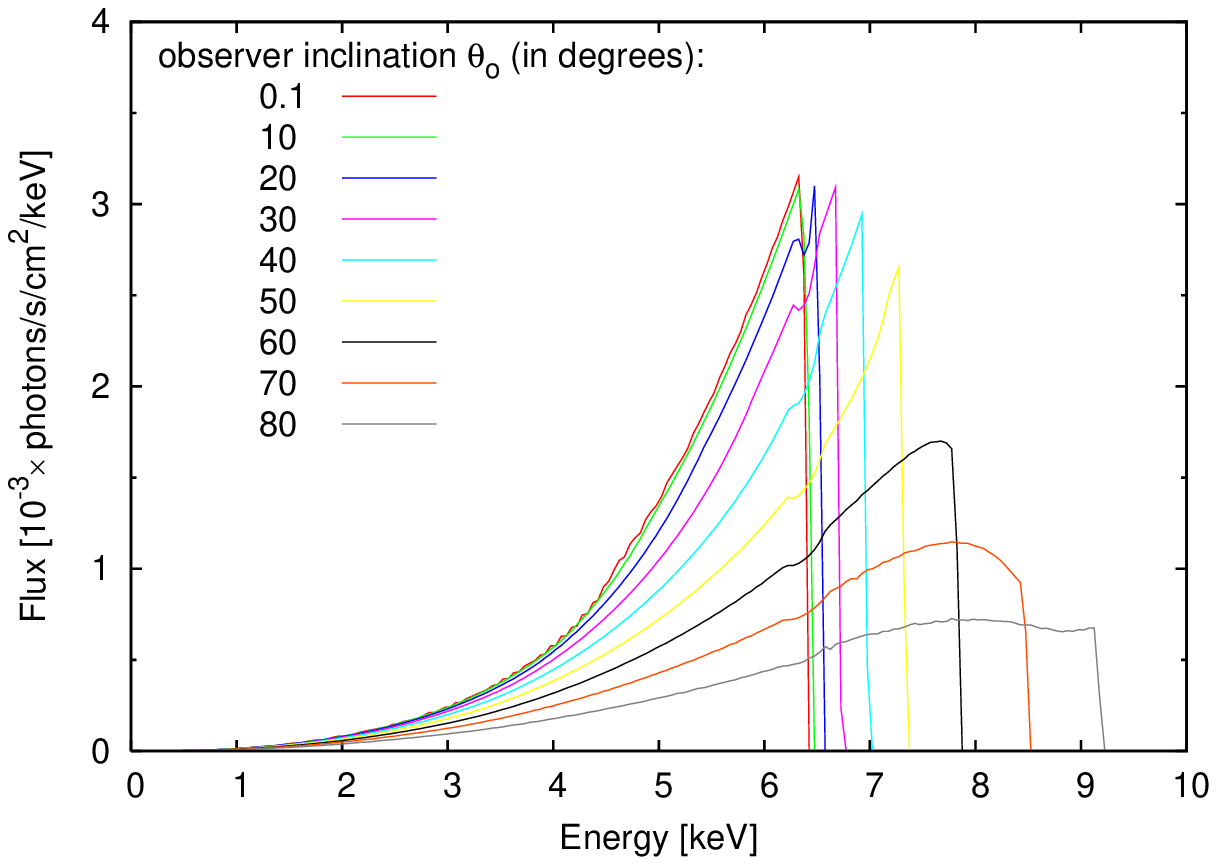}{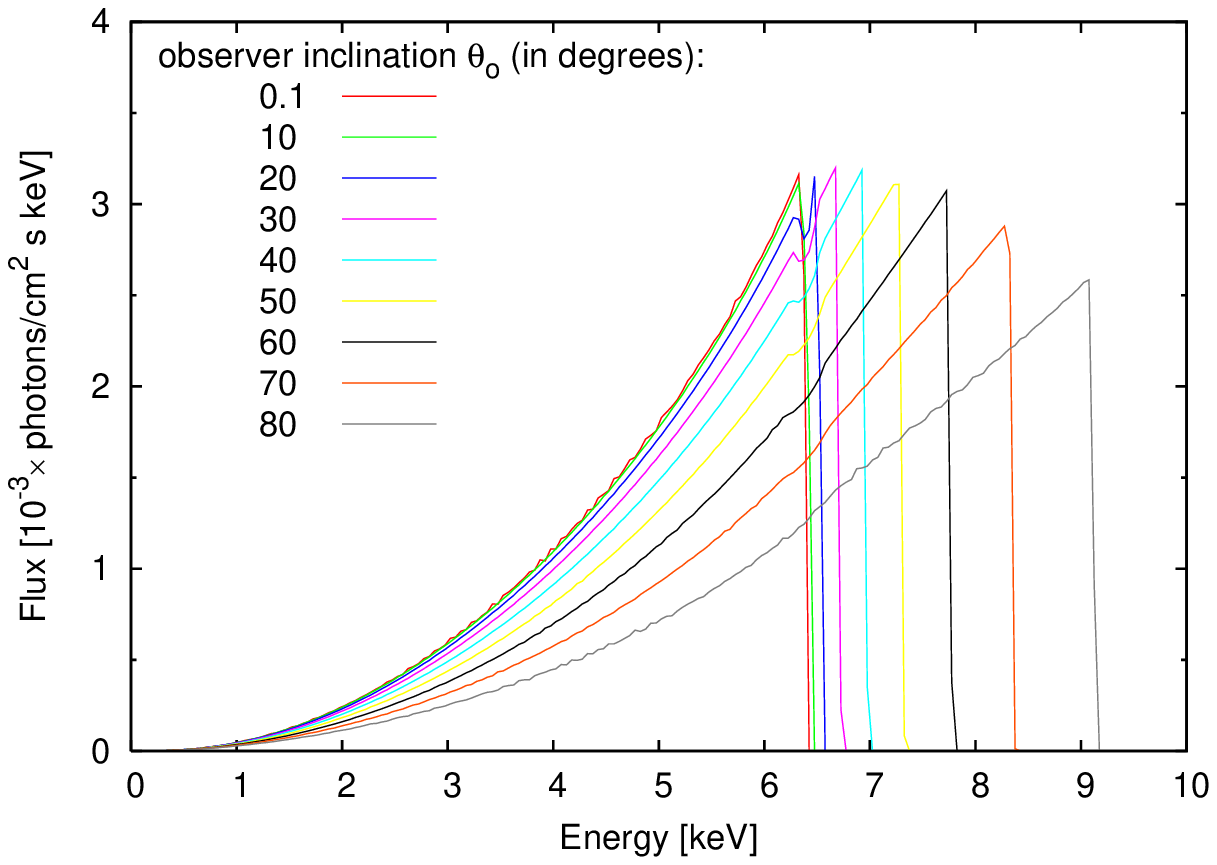}
\caption{Dependence on observer inclination (values of $\theta_{\rm{}o}$ are 
given in the plot). Left: non-rotating black hole, $a=0$.
Right: maximally rotating black hole, $a=1$.
Other parameters as in Fig.~\ref{fig:example0}.}
\label{fig:example3}
\end{figure*}

Geometrical optics is adequate to formulate the ray tracing
problem in rigorous manner (Schneider, Ehlers \& Falco 1992).
Physically, one has to consider a ray bundle connecting the source
and the observer. We realize that the photon number is conserved 
in a bundle along the photon ray. Furthermore, this number is
observer-independent quantity. Four-momentum of each photon is 
parallely transported along the ray. On a practical side, to achieve
fast computations of observed spectra and for data fitting we
produced accurate tables describing the Kerr metric, ray-tracing
equations and the equations of geodesic deviation. These were solved in 
well-behaved coordinates. This choice allows one
to map the whole region, from the outer edge of the disk down to the
black hole horizon, where dragging effects are prominent. Also, caustics
can play a role for large inclination (Rauch \& Blandford 1994). 
When integration of light rays is accomplished, we transform results
back to Boyer-Lindquist coordinates, so that the intrinsic emissivity of
the disc can be conveniently defined as a function of $r$, $\varphi$
and $t$ in the disk plane. Here we briefly summarise the adopted approach.

The source appears as a point-like object for a distant observer,
so that the observer actually measures the flux entering the solid 
angle ${\rm d}\Omega_{\rm obs}$, which is associated with the detector area 
${\rm d}S_{\rm obs}{\equiv}D^2\,{\rm d}\Omega_{\rm obs}$ (this relation 
defines the distance $D$). We denote total photon flux received by a 
detector, 
\begin{equation}
\label{flux}
N^{S}_{\rm{}obs}(E)\equiv\frac{{\rm d}n(E)}{{\rm d}t\,{\rm d}S_{\rm obs}}
={\int}N_{\rm loc}(E/g)\,g^2\,{\rm d}\Omega,
\end{equation}
where  
\begin{equation}
N_{\rm loc}(E_{\rm loc})\equiv
\frac{{\rm d}n_{\rm loc}(E_{\rm loc})}{{\rm d}\tau\,{\rm d}S_{\rm loc}\,
{\rm d}\Omega_{\rm loc}}
\end{equation}
is local photon flux at the disk, ${\rm d}n(E)$ is the number of photons
with energy in interval ${\langle}E,E+{\rm d}E\rangle$ and 
$g=E/E_{\rm{}loc}$ is the redshift factor. Note that
$N_{\rm{}loc}(E_{\rm{}loc})$ may vary over the disk, change in time, and
it can also depend on the emission angle (for the sake of brevity of
formulae, we do not always write this dependence explicitly).

The emission arriving in solid angle ${{\rm d}\Omega}$ originates
from area ${\rm d}S_{\rm loc}$ on the disk. Hence, in our computations
we want to integrate the flux contributions over a fine mesh on the disk
surface. To this aim, we adjust eq.~(\ref{flux}) to the form
\begin{equation}
N^S_{\rm{}obs}(E)\equiv
 \frac{1}{D^2}{\int}N_{\rm loc}(E/g)\,g^2\,\frac{D^2\,{\rm d}
 \Omega}{{\rm d}S_{\rm loc}}\,{\rm d}S_{\rm loc}
=\frac{1}{D^2}{\int}N_{\rm{}loc}(E/g)\,g^2\,
 \frac{{\rm d}S_{\rm f}}{{\rm d}S_{\rm loc}}
 \,\frac{{\rm{}d}S_{\rm loc}}{{\rm d}S}\,{\rm d}S\,.
\end{equation}
Here ${{\rm{}d}S_{\rm f}}$ denotes an element of area on the disk
perpendicular to light rays (corresponding to the solid angle ${\rm
d}\Omega$ in flat space-time), and ${\rm d}S\equiv r\,{\rm d}r\,{\rm
d}\varphi$ stands for area in Boyer-Lindquist coordinates lying in the
disk plane from which photons are emitted.
The integrated flux per unit solid angle is
\begin{eqnarray}
{\rm d}N^{\Omega}_{\rm obs}(E,t) & \equiv &
 N^{\Omega}_{\rm obs}(E,t)\,{\rm d}E
\nonumber \\
 & = & N_0\int_{r_{\rm in}}^{r_{\rm out}} \!\! {\rm d}r
 \int_{\varphi_{\rm{}min}}^{\varphi_{\rm{}max}} \!\!{\rm d}\varphi
 \int_{E/g}^{(E+{\rm d}E)/g} \!\!{\rm d}E_{\rm loc}\,
 N_{\rm loc}(E_{\rm loc},r,\varphi,\mu_{\rm loc},t_0)\,g^2\,l\,r,
\label{nobs}
\end{eqnarray}
where $t_0\,\equiv\,t-\Delta t$ is time coordinate corrected for 
the light-time effect,
\begin{equation}
l=\frac{{\rm d}S_{\rm f}}{{\rm{}d}S_{\rm loc}}=
\frac{{\rm d}S_{\rm f}}{{\rm d}S_\perp}\mu_{\rm loc}
\end{equation} 
is the lensing factor in the limit $D\rightarrow\infty$ 
(keeping $D^2\,{\rm d}\Omega$ constant), $\mu_{\rm loc}$ 
is the cosine of the local emission angle,
${\rm d}S_{\perp}$ is proper area at the
disk perpendicular to light rays and corresponding to
${\rm d}\Omega$. In eq.~(\ref{nobs}) we used 
transformation of the proper area element, ${{\rm d} S_{\rm loc}}$,
\begin{equation}
\frac{{\rm d} S_{\rm loc}}{{\rm d} S}=g^{-1}.
\end{equation} 
Finally, $N^{\Omega}_{\rm obs}(E,t)$ is the output of the model
computations.

In eq.~(\ref{nobs}),$N_0$ is a normalization constant of the model. For
line emission, normalization is chosen in such a way that the total flux
from the disk is unity. In the case of a continuum model, the flux is
normalized to unity  at certain value of observed energy (typically at
$E=1$~keV, as in other {\sc{}xspec} models).  Notice that default
normalization has been changed in  Figs.\
\ref{fig:example0}--\ref{fig:example3} for the sake of clarity.

We employed the Bulirsch-Stoer method for numerical integration.
Four sets of data tables were pre-computed and stored with the necessary
information about photons received from different regions of the source.
Specifically, separate tables hold information on
(i)~the energy shift ($g$-factor, or the ratio of energy of 
a~photon received by an observer at infinity to the intrinsic 
energy when emitted at the disk), (ii)~the lensing effect (the ratio $l$ of the 
area subtended by photons at infinity, perpendicular to light rays 
through which photons arrive at the detector, to the area on the disk from
where these photons originate), (iii)~the mutual delay of the photons 
arriving  at the observer's location (i.e.\ relative time lags $\Delta{t}$), 
and (iv)~the local emission direction $\mu_{\rm{}loc}$ with respect to the 
disk normal (evaluated in the disk co-rotating frame). 

The problem of directional distribution of the reflected radiation is
quite complicated and the matter has not been completely  settled yet
(e.g.\ George \& Fabian 1991; Ghisellini, Haardt \& Matt 1994; 
\.{Z}ycki \& Czerny 1994;  Magdziarz \& Zdiarski 1995). A specific angular
dependence is often assumed in  models, such as the limb-darkening in
{\sf{}laor} {\sc xspec} model,  $I(\mu_{\rm
loc})\,\propto\,1+2.06\mu_{\rm loc}$.  However, it has been argued that
limb-brightening may  actually occur in the case of strong primary
irradiation of the disk, and this is relevant for accretion disks near
black holes where the effects of emission anisotropy are crucial (Czerny
et al.\ 2004). We find that the spectrum of the inner disk turns out to
be very sensitive to the adopted angular dependence of the emission, and
so the possibility to modify this profile and examine the results using
{\sf{}ky} appears to be rather useful (we set locally isotropic emission
as default). Impact of the darkening law on the reflection spectrum is
connected with the assumed rotation law of the disk material. As
mentioned above,  we assumed Keplerian circular motion for
$r{\geq}r_{\rm{}ms}$. For $r<r_{\rm{}ms}$ we assumed free-fall
trajectories, maintaining constant orbital energy and angular momentum.

All data tables are 
stored in {\sc{}fits} format\footnote{Flexible Image Transport 
System. See e.g.\ Hanisch et al.\ (2001) or 
{\sf{}\url{http://fits.gsfc.nasa.gov/}} for specifications.}
and loaded in memory only once, when the routine is initiated. 
Examples of the contents of the data tables, illustrating the 
effects of frame-dragging, are shown in 
Figures~\ref{fig:example6a}--\ref{fig:example6b} for  
given choices of $a$ and $\theta_{\rm{}o}$.
The graphs represent a top view of the equatorial plane, each one on
four different spatial scales.
In practice we have chosen $r_{\rm{}out}=10^3$ 
(Boyer-Lindquist radius) as a maximum outer limit of the disk,
but the most dramatic dependences on, for example $a$,
happen much closer to the horizon. 
In the upper left panels of Figs.~\ref{fig:example6a} and
\ref{fig:example6b} we define the radial coordinate 
in a different way to that in the other panels. Specifically, we use
${r^{\prime}}^2\equiv{x^{\prime}}^2+{y^{\prime}}^2=(r-r_{\rm h})^2$.
The use of $r^{\prime}$ brings the horizon to the origin
so that the region 
just outside it is well resolved in the plots that show the region
closest to the black hole. In these graphs, the
observer is located at the top of the pictures with 
an inclination angle of $30^{\circ}$ relative to the rotation axis.
The marginally stable orbit is also shown (drawn as a circle) where 
relevant. 

Despite the fact that the particular choice of the graphical 
representation is a technicality to certain extent, we find these kinds 
of graphs useful for quick estimations of the expected range of energy
shifts, fluctuations of radiation flux, time delay effects etc. For
example, we notice that with the adopted (typical) value of
$\theta_{\rm{}o}=30^{\circ}$, the lensing cannot account for more than a
few percent effect on observed count rates. Lensing, however, becomes
important for  edge-on orientation of the disk, as can be seen from the
tables with larger values of $\theta_{\rm{}o}$. We therefore arranged a
systematic atlas of the tables for different values of angular momentum 
and observer inclination. Further, we produced their transformation 
from Boyer-Lindquist $(r,\varphi,t)$ to Kerr ingoing coordinates (in
which the effect of frame dragging is largely eliminated), and the same
for other relevant quantities (time delay and the emission angle in the 
disk). It is worth noticing that different space-time metric and the
disk rotation law can be accomodated simply by replacing the data
tables. Also, transfer  of Stokes parameters can be computed, and to
this aim we computed data tables which are needed for polarimetry in
strong gravity. However,  this discussion remains beyond the scope of
the present paper.\footnote{The  corresponding graphs can be obtained
from an internet site, {\sf{}\url{http://astro.mff.cuni.cz/ky/}}.} 

\begin{figure*}[tbh]
\epsscale{1}
\plottwo{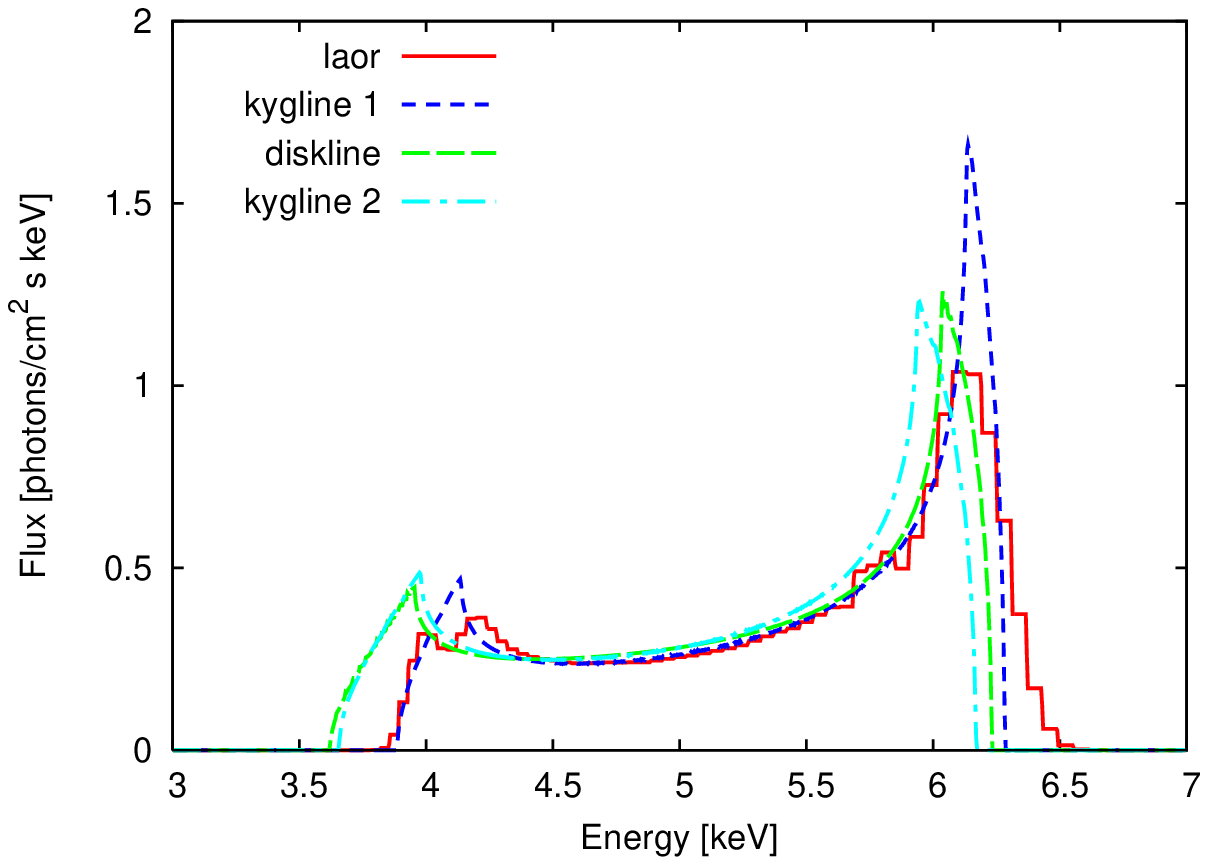}{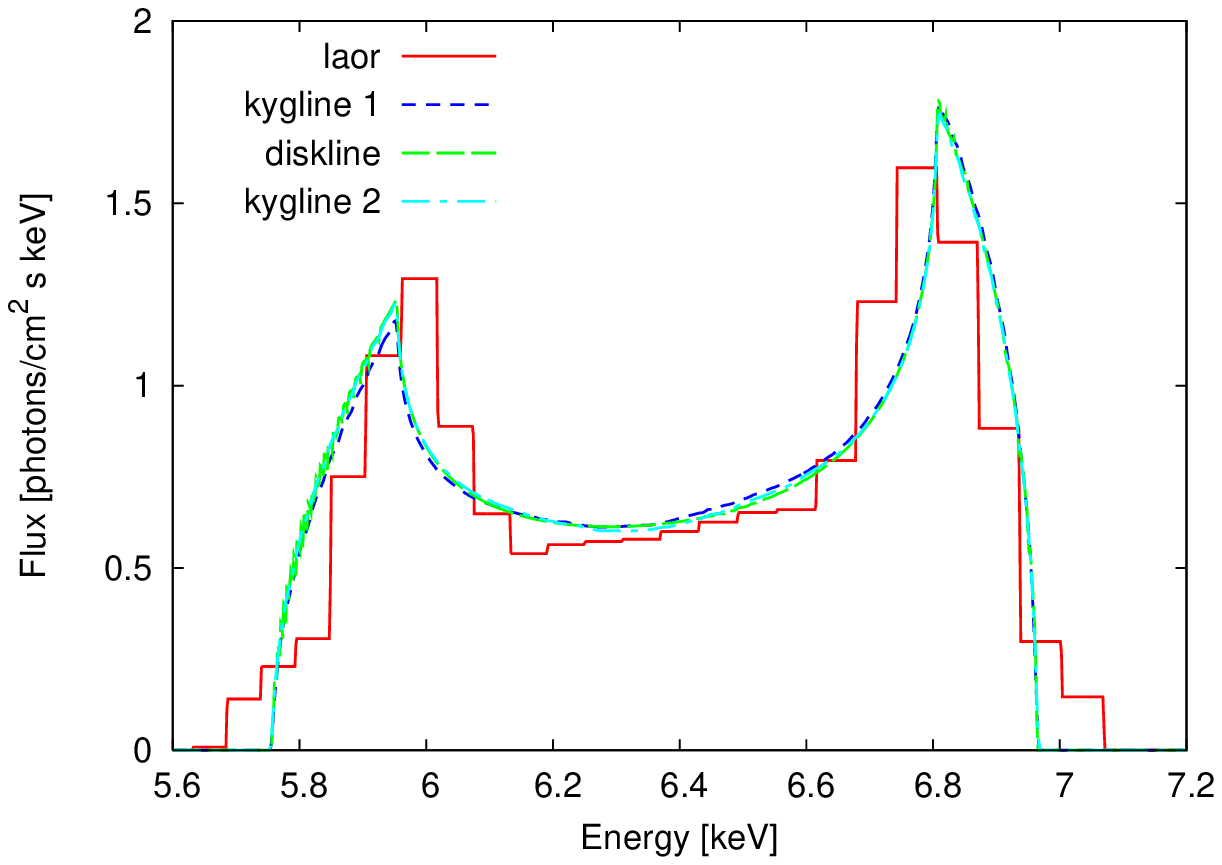}
\caption{Comparison of the output from {\sc{}xspec} models
for the disk-line problem: {\sf{}laor}, {\sf{}diskline}, and
{\sf{}kygline} (line 1 corresponds to the same limb-darkening 
law and $a=0.9982$ as in {\sf{}laor}; line 2 assumes locally 
isotropic emission and $a=0$ as in {\sf{}diskline}). Left panel: 
$\theta_{\rm{}o}=30^{\circ}$, $r_{\rm{}in}=6$, $r_{\rm{}out}=7$. 
Right panel: $\theta_{\rm{}o}=70^{\circ}$, $r_{\rm{}in}=100$, $r_{\rm{}out}=200$.
Radial decay of intrinsic emissivity follows $\alpha=3$ power law.
Default normalization of the model has been retained in this plot.}
\label{fig:example4}
\end{figure*}

\begin{figure*}[tbh]
\epsscale{1}
\plottwo{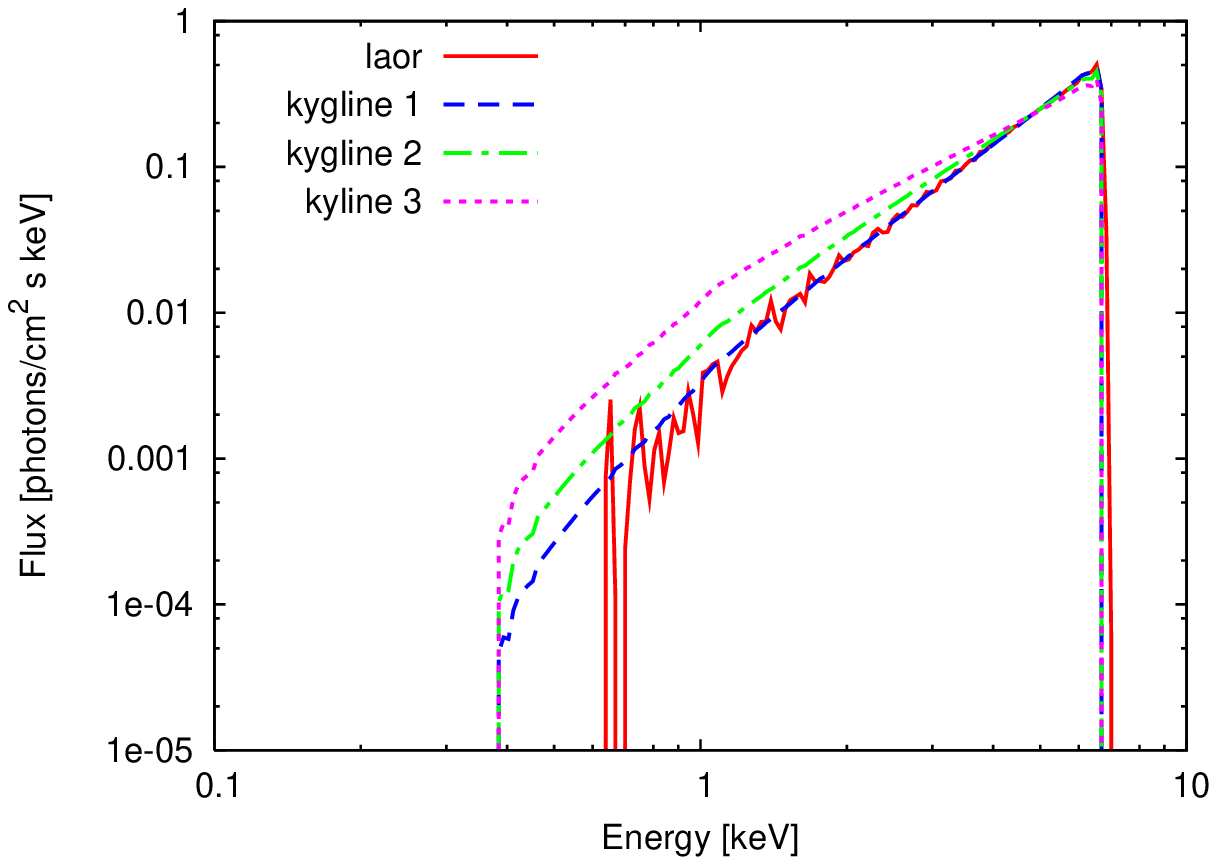}{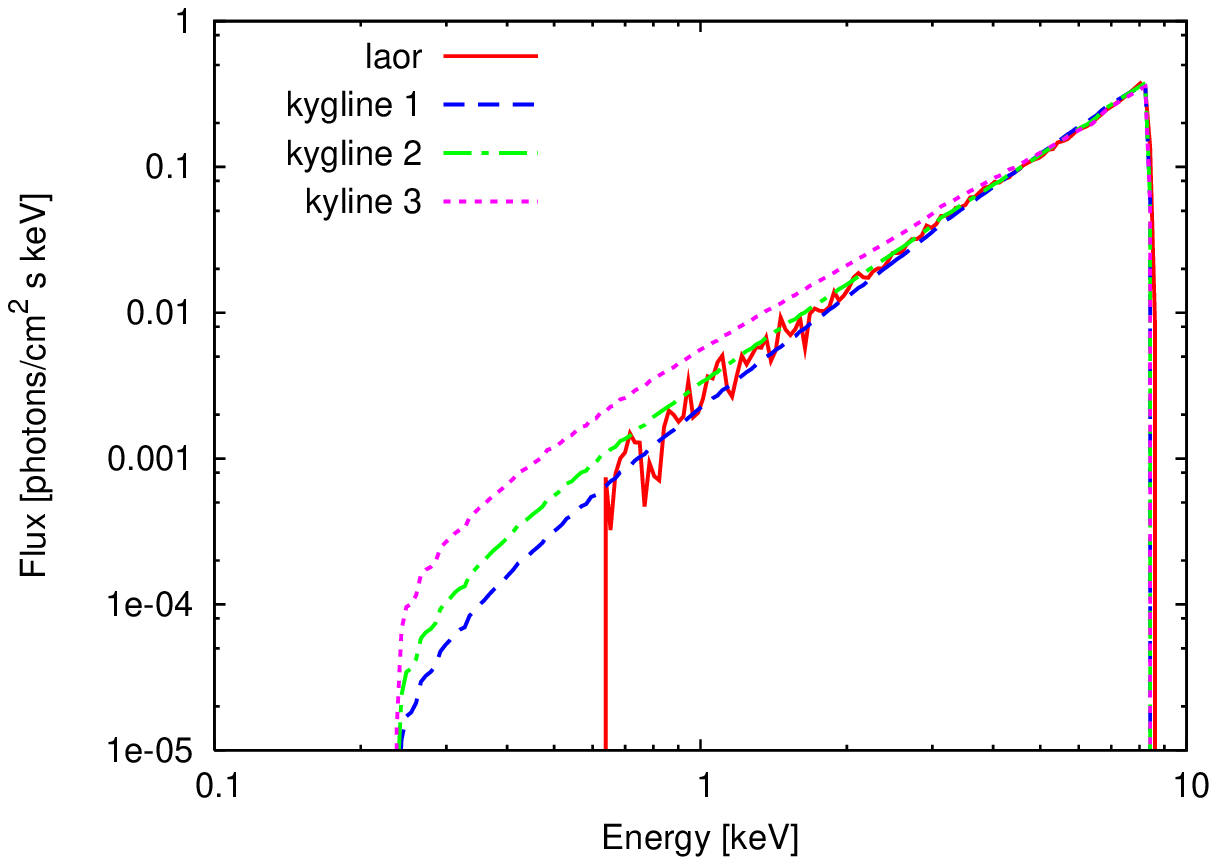}
\caption{Comparison between the {\sf{}laor} and {\sf{}kygline}
models. These plots are similar to previous figure, but in logarithmic 
scale and for three choices of the darkening law in {\sf{}kygline} 
--- (1)~$I\propto1+2.06\mu_{\rm{}loc}$; (2)~$I={\rm{}const}$;
(3) $I\propto\log(1+1/\mu_{\rm{}loc})$. Notice the impact on
the calculated red-wing slope.
Left panel: $\theta_{\rm{}o}=30^{\circ}$; 
Right panel:  $\theta_{\rm{}o}=70^{\circ}$. In both panels, 
$r_{\rm{}in}=r_{\rm{}ms}$, $r_{\rm{}out}=400$, $a=0.9982$.}
\label{fig:example5}
\end{figure*}

Results of an elementary code test are shown in
Figures~\ref{fig:example0}--\ref{fig:example3}. The intrinsic emissivity
was assumed to be a narrow gaussian line (width
$\sigma\dot{=}0.42\,\mbox{FWHM}=5$\,eV) with the amplitude decreasing
$\propto{}r^{-\alpha}$ in the local frame co-moving with the disk
medium. No background continuum is included here, so these lines can be
compared with similar pure disk-line profiles obtained in previous
papers (e.g.\ Laor 1991; Kojima 1991) which also imposed the assumption of
axially symmetric and steady emission from an irradiated thin disk.
Again, the intrinsic width of the line is assumed to be much less than
the effects of broadening due to bulk Keplerian motion and the central
gravitational field. Furthermore,
Figures~\ref{fig:example4}--\ref{fig:example5} compare model spectra 
of widely used {\sc{}xspec} models.
 
Typically, the slope of the low-energy wing is rather sensitive do the
radial dependence of emissivity. Notice also
that the {\sf{}laor} model gives zero contribution at energy below
$0.1E_{\rm{}loc}=0.64\,$keV and
its grid has only $35$ radial points distributed in the whole range
$1.23{\leq}r{\leq}400$. Therefore, in spite of very efficient
interpolation and smoothing of the final spectrum, the {\sf{}laor} model
does not accurately reproduce the line originating from a narrow ring.
Also, dependence on the limb darkening/brightening cannot be examined
with this model, because the form of directionality of the intrinsic
emission is hard-wired in the code, together with the position of the
inner edge at $r{\geq}r_{\rm{}ms}$. 

The {\sf{}diskline} model has been also used frequently in the context
of spectral fitting, assuming a disk around a non-rotating black hole.
This model is analytical, and so it has clear advantages in
{\sc{}xspec}. Notice, however, that the lensing effect is neglected. All
this affects especially the spectrum near the black hole, where
radiation is expected to be very anisotropic and flow lines
non-circular.

\section{Fast Model for Compton Reflection}
\label{appb}
In this Appendix we describe the Compton-reflection model {\sf{}hrefl}. 
This model has been a part of the {\sc{}xspec} distribution
for some time. We have used similar approximation for the
Compton-reflection in the disk to produce the relativistic 
version, {\sf{}kyhrefl}.

The model has similar features to the well-known {\sf{}pexrav} 
model (Magdziarz \& Zdziarski 1995), which can often
be prohibitively slow. There is a need for a fast reflection
code for spectral fitting purposes, since the
energy resolution of X-ray spectral data has been improved considerably
in the last few years. Also, when convolving the reflection continuum with a
kernel for relativistic energy shifting, computation speed
is a critical factor. 
The price to pay is that {\sf{}hrefl} makes use of a number
of simplifying approximations. One of these
is the elastic scattering approximation. Still, {\sf{}hrefl} is
accurate enough to be used for
modelling data from observations of low-redshift 
sources with {\it ASCA}, {\it Chandra}, and {\it{}XMM--Newton},
where the source-frame spectrum does not extend beyond $\sim15$~keV.
(Notice, however, that this condition may be violated if the
redshift of received photons is very large, e.g.\ near a black-hole horizon.)
Another advantage of {\sf{}hrefl} is that the model is completely
analytic, and this is one of the major factors which makes it fast. 

Below we give an outline of the model, providing important details of
the adopted approximations, and a direct comparison with the
{\sf{}pexrav} model.  A general description of the approach can be found
in Basko  (1978) along with any standard radiative transfer textbooks
(e.g.\ Chandrasekhar 1960).

\begin{figure*}[tbh]
\epsscale{1}
\plottwo{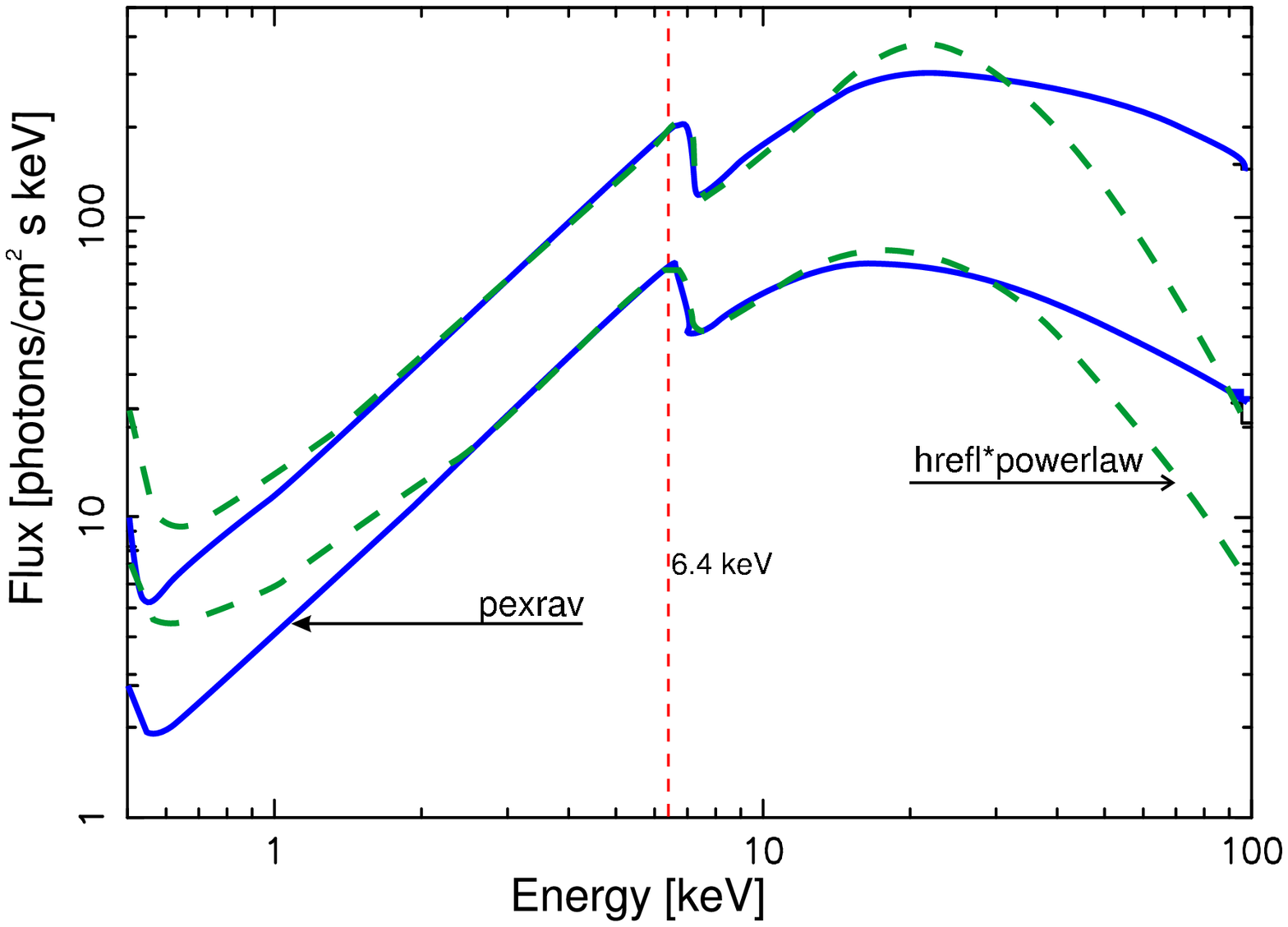}{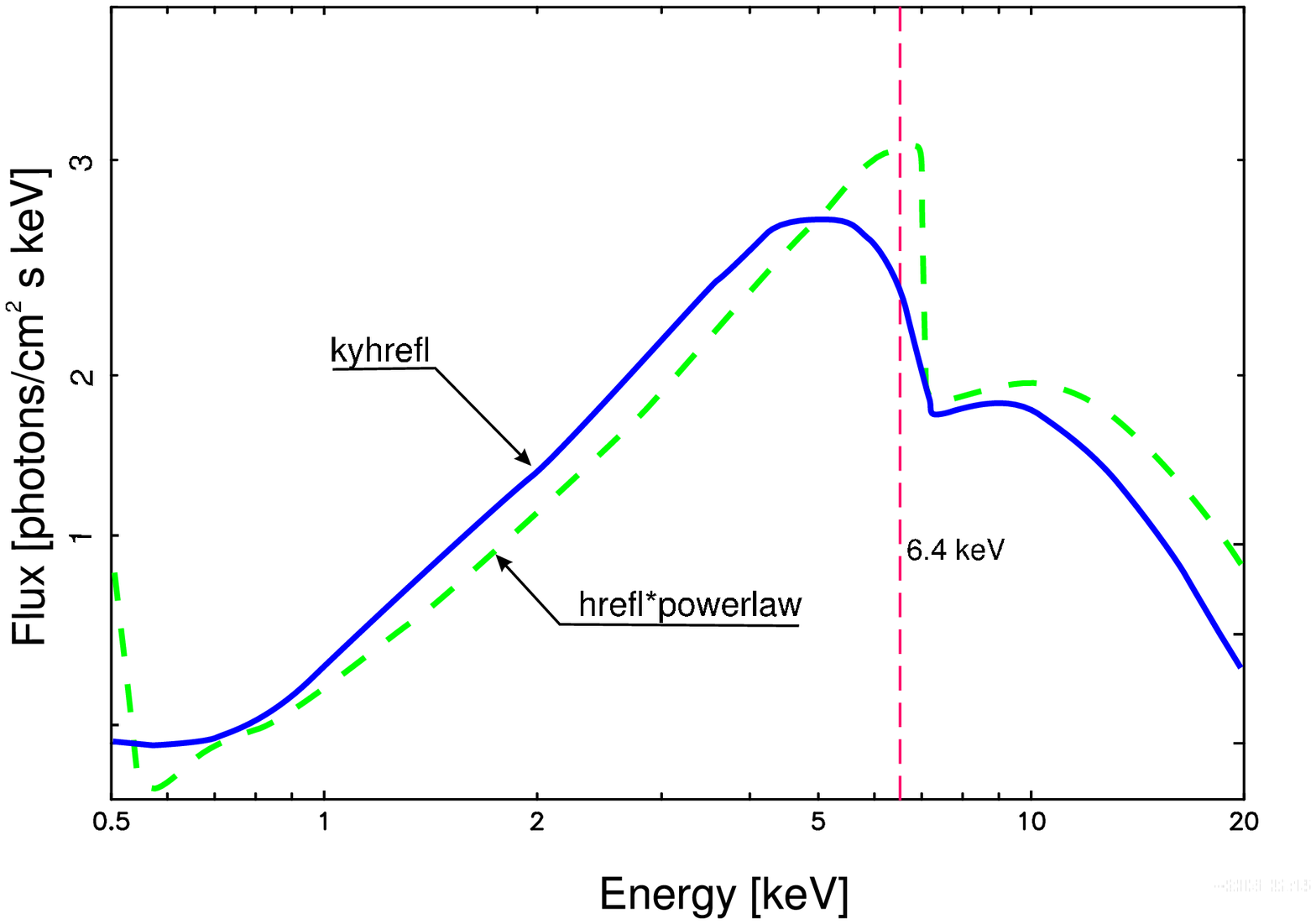}
\caption{Left: Comparison between {\sf{}pexrav} (solid) and
{\sf{}hrefl*powerlaw} (dashed) for two different inclination angles.
Approximations are involved, making the
model valid below $E\sim15$~keV in the source rest-frame.
Right: Comparison between the models {\sf{}kyhrefl}
(solid line) and {\sf{}hrefl*powerlaw} (dashed line). Only the 
reflection component of the radiation flux due to the primary power-law 
continuum is plotted (in arbitrary units).
Default model parameters were used with $\theta_{\rm{}o}=30^{\circ}$.
\label{fig:example8}}
\end{figure*}

Let us consider a point source situated at a height
$h$ above the center of an optically-thick disk with inner and outer
radii $r_{\rm{}in}$ and $r_{\rm{}out}$ respectively. 
We define the cosines of the
angles between the disk axis
and lines connecting the inner and outer radii to the source as
$\mu_{\rm min}$ and $\mu_{\rm max}$ respectively, where
\begin{equation}
\mu_{\rm min} = \frac{h}{\sqrt{r_{\rm{}in}^{2}+h^{2}}}
\quad\mbox{and} \quad
\mu_{\rm max} = \frac{h}{\sqrt{r_{\rm{}out}^{2}+h^{2}}}.
\end{equation}
The parameters actually required by the {\sf{}hrefl} model are
$\theta_{\rm min} = \arccos{\mu_{\rm min}}$ and
$\theta_{\rm max} = \arccos{\mu_{\rm max}}$.
In the approximation of an infinitely large disk with no hole at the
center, $\theta_{\rm min}\rightarrow0$ and
$\theta_{\rm max}\rightarrow\pi/2$. 

Now we adopt cylindrical coordinates with
an origin coincident with the center of the disk face,
and a set of rays emanating from the source and incident on
the disk, making an angle 
$\arccos{\mu_{\rm{}i}} = \arccos{[h/\sqrt{h^{2}+r^{2}}]}$ 
with the disk normal at the point of interception.
The incident specific intensity is $I_{\rm{}i}(r,\varphi)$. 
In the case of diffusive reflection from a plane-parallel atmosphere,
the standard solution for the specific intensity of the reflected rays
(as a function of energy, $\epsilon$) gives

\begin{equation}
I(\mu_{\rm{}r}, \epsilon) =
 \frac{\lambda(\epsilon)}{4\pi}\;I_{\rm{}i}\,\mu_{\rm{}i}\; 
 \frac{H[\mu_{\rm{}i},\lambda(\epsilon)] \,
 H[\mu_{\rm{}r},\lambda(\epsilon)]}{\mu_{\rm{}i} + \mu_{\rm{}r}} 
 \quad[{\rm  keV \ cm^{-2} \ s^{-1} \ sr^{-1} \ keV^{-1}}].
\label{eq:app3}
\end{equation}
Here, $\mu_{\rm{}r}$ is the cosine of the reflected rays with respect to
the disk normal, $\lambda(\epsilon)$ is the single-scattering albedo
(i.e.\ ratio of Thomson-scattering cross-section to the total
scattering plus absorption cross-section), and $H[\mu, \lambda(\epsilon)]$
are the Chandrasekhar $H$-functions for isotropic scattering.
The angles are to be evaluated in the frame co-rotating with the disk
medium, which is particularly important in the relativistic version
of the code when the bulk orbital motion is very rapid close to
the black hole. The assumption of isotropic scattering is one of
the simplifications we employ: at photon energies much less than
the electron rest-mass, the angular dependence of photon-electron
scattering is not very strong. The albedo is defined as

\begin{equation}
 \lambda(\epsilon) =
 \frac{1.2\,\sigma_{\rm{T}}}{1.2\,
 \sigma_{\rm{T}}+
 \sigma_{\rm{abs}}(\epsilon)},
\end{equation}
where $\sigma_{\rm{T}}$ is the Thomson cross-section 
and $\sigma_{\rm{abs}}$
is the photoelectric absorption cross-section. The numerical 
factor $1.2$ accounts for 
electrons from ionized helium (those from heavier elements are
neglected due to their smaller numbers). 

Now we can reformulate equation (\ref{eq:app3}) in terms of actual fluxes.
Let $N_{\rm{}d}(\epsilon)$ be the {\em{}direct\/} spectrum 
(${\rm photons \ cm^{-2} \ s^{-1} \ keV^{-1}}$) of an
X-ray source, observed at a large distance $D$.
Further, ${\rm{}d}F_{\rm{}i}$ denotes the energy flux
within an energy interval ${\rm{}d}\epsilon$, which is 
incident on the face of the disk at radius $r$ from the disk center:

\begin{equation}
{\rm{}d}F_{\rm{}i} = I_{\rm{}i}\, \mu_{\rm{}i}  = 
\frac{4 \pi D^{2} \epsilon\,N_{\rm{}d}(\epsilon)\,{\rm{}d}\epsilon}
{4\pi (r^{2}+h^{2})}\; \mu_{\rm{}i} \quad[{\rm{}keV \ cm^{-2} \ s^{-1}}].
\label{eq:app5}
\end{equation}
The reflected flux that is received by a detector 
having an effective area $A$
from a disk surface element of area, 
$r\,{\rm{}d}r\,{\rm{}d}\varphi$, is then

\begin{equation}
{\rm{}d}F_{\rm{}r} = {\mu_{\rm{}r{}}} 
I({\mu_{\rm{}r{}}},\epsilon)\, r \,{\rm{}d}r\,{\rm{}d}\varphi \ \ 4\pi 
\left(\frac{A}{4\pi D^{2}}\right)
\frac{1}{A} \quad[{\rm keV \ cm^{-2} \ s^{-1}}].
\label{eq:app6}
\end{equation}
Here, $4\pi (A/4\pi D^{2})$ is the solid angle subtended by the
detector, $r=h\tan{\theta_{\rm{}i}}$, and
${\rm{}d}r = h\,{\rm{}d}\theta_{\rm{}i}/\mu_{\rm{}i}^{2}$.

Further, we denote the observed reflected spectrum 
$N_{\rm{}r}(\epsilon)$ $[{\rm photons \ \rm cm^{-2} \ s^{-1} \ keV^{-1}}]$. 
Since ${\rm{}d}F_{\rm{}r} = \epsilon\,{\rm{}d}N_{\rm{}r}(\epsilon) 
\,{\rm{}d}\epsilon\, {\rm{}d}{\mu_{\rm{}r{}}}$, we can 
substitute eq.~(\ref{eq:app5}) into eq.~(\ref{eq:app3}), then
eq.~(\ref{eq:app3}) into eq.~(\ref{eq:app6}), and by integration 
over $\varphi$ from $0$ to $2\pi$ we find

\begin{equation}
 N_{\rm{}r}(\epsilon) = 
 {\textstyle{\frac{1}{2}}}\lambda(\epsilon)\, N_{\rm{}d}(\epsilon)
 \int_{\mu_{\rm max}}^{\mu_{\rm min}}
 {\frac{{\mu_{\rm{}r{}}} \,{\rm{}d}\mu_{\rm{}i}}{\mu_{\rm{}i} + {\mu_{\rm{}r{}}}} }\, 
 H[\mu_{\rm{}i},\lambda(\epsilon)]\, H[\mu_{\rm{}r},\lambda(\epsilon)].
\label{eq:app8}
\end{equation}
For the $H$-functions we use an approximation (Basko 1978)
which is good to $\sim 8\%$ accuracy in the range $0 \le \lambda(\epsilon)
\le 1$ and $0 \le \mu \le 1$:
\begin{equation}
H[\mu,\lambda(\epsilon)] \sim \frac{1 + \sqrt{3}\, \mu}
{1 + \sqrt{3\left(1-\lambda(\epsilon)\right)}\; \mu}.
\end{equation}
Next, we notice that for a given $\lambda(\epsilon)$, 
$H[\mu_{\rm{}i},\lambda(\epsilon)]$
does not vary much, so we can bring it in front of the integral, 
replacing the exact formula with
the mean over the range $\mu_{\rm{}i}=0$ to $\mu_{\rm{}i}=1$,
\begin{equation}
\bar{H}[\lambda(\epsilon)] = 
\int_{0}^{1}{H[\mu_{\rm{}i},\lambda(\epsilon)]\,{\rm{}d}\mu_{\rm{}i}}.
\end{equation}
For $\lambda(\epsilon)<1$ we obtain
\begin{equation}
\bar{H}[\lambda(\epsilon)] = \frac{1-\sqrt{1-\lambda(\epsilon)}}
{\sqrt{3}\left(1-\lambda(\epsilon)\right)} \;
\ln{\left[1+\sqrt{3\left(1-\lambda(\epsilon)\right)}\right]}
+\frac{1}{\sqrt{1-\lambda(\epsilon)}},
\end{equation}
and for $\lambda(\epsilon)=1$ 
\begin{equation}
\bar{H}[\lambda(\epsilon)] = 1 + {\textstyle{\frac{1}{2}}}\,\sqrt{3}.
\end{equation}
Finally, we integrate equation (\ref{eq:app8}) over $\mu_{\rm{}i}$ to find
\begin{equation}
 N_{\rm{}r}(\epsilon,\mu_{\rm{}r{}}) = {\textstyle{\frac{1}{2}}}\, 
 N_{\rm{}d}(\epsilon,\mu_{\rm{}r{}})\, \lambda(\epsilon)\, \mu_{\rm{}r{}}\, 
 \ln{\left(\frac{\mu_{\rm{}r{}} + \mu_{\rm min}}{\mu_{\rm{}r{}} + \mu_{\rm max}}\right)}
 \bar{H}[\lambda(\epsilon)]\, H[\mu_{\rm{}r{}},\lambda(\epsilon)]. 
\end{equation}
The function {\sf{}hrefl} is encoded as a multiplicative component 
in {\sc{}xspec}. When multiplied by the direct observed continuum,
{\sf{}hrefl} returns the reflected continuum, which is further multiplied by 
an effective reflection fraction, $R_{\rm{}c}$. The resulting contribution is 
added to the fraction $R_{\rm{}d}$ of direct continuum flux. 
In other words, $R_{\rm{}c}=1$ corresponds to $2\pi$ of the disk covering
solid angle as seen by the X-ray source. Thus,

\begin{equation}
N_{\rm{}d}(\epsilon,\mu_{\rm{}r{}})\,\times\,
\mbox{\sf{}hrefl}(\theta_{\rm min}, \theta_{\rm max},
\theta_{\rm{}o}, R_{\rm{}c}, R_{\rm{}d}, \epsilon) 
= N_{\rm{}d}(\epsilon,\mu_{\rm{}r{}}) \left[R_{\rm{}d} + R_{\rm{}c} 
\frac{N_{\rm{}r}(\epsilon,\mu_{\rm{}r{}})}{N_{\rm{}d}(\epsilon,\mu_{\rm{}r{}})} \right],
\end{equation}
where $\mu_{\rm{}r{}}$ has the meaning of observer inclination.
The arguments of {\sf{}hrefl} are the free parameters of the model
(there are two more that are not listed explicitly here:
they are the iron abundance relative to solar, and energy of the Fe-K edge).
Notice that $R_{\rm{}c}$ and $R_{\rm{}d}$ are {\em{}not independent}, so 
no more than one of these two parameters is allowed to be free
at any time of the fitting procedure. This way of coding the 
model is simply a matter of convenience for the user.

Fig.~\ref{fig:example8} (left panel) compares the pure reflection spectrum 
produced by {\sf{}hrefl} with the {\sf{}pexrav} model using an input power law
(photon index $\Gamma=2$) for two inclination angles 
($\mu_{\rm{}r} = 0.95$ and $\mu_{\rm{}r} =0.1$, corresponding to 
$\theta_{\rm{}o} = 18.2^{\circ}$ and
$84.3^{\circ}$ respectively; {\sf{}pexrav} does not allow for values of
$\mu_{\rm{}r}<0.95$). In {\sf{}hrefl}
the parameters $\theta_{\rm min}$ and $\theta_{\rm max}$ were
fixed at $0^{\circ}$ and $90^{\circ}$ respectively for the spectral 
fitting described in the present paper. 
Note that {\it the relative iron abundance parameter\/}
for {\sf{}hrefl} was set to $1.418=4.68 \times 10^{-5}/3.3 \times 10^{-5}$
because the two codes use different iron abundances. 

Apart from the adopted approximations, additional discrepancies
arise between {\sf{}hrefl} and {\sf{}pexrav} due
to the use of different photoelectric absorption cross-sections
({\sf{}hrefl} uses older versions of cross-section values than
{\sf{}pexrav} and
the difference is visible at low energies of the spectrum).
Between 1--15~keV the errors are less than 20\% (much better accuracy 
than this value is achieved for the face-on inclinations). 
If the reflection continuum is combined
with the direct continuum, for typical covering factors ($R_{\rm{}c}$), the errors
below the Fe-K edge are practically negligible.

We conclude with the following concise summary of the approximations used in
{\sf{}hrefl}. The starting point of the model is the standard solution
of the radiative transfer equation for reflection in a plane-parallel
atmosphere. The simplifying assumptions are as follows.
(i)~Isotropic elastic scattering is assumed.
(ii)~Approximate analytical forms of the Chandrasekhar $H$-functions
are used, as described in Basko (1978), but the accuracy of the approximate 
formulae is better than $\sim 8\%$.
(iii)~When integrating the incident continuum flux on the disk
surface, the corresponding $H$-function is taken outside
the integral and replaced with its mean value, $\bar{H}$.

Fig.~\ref{fig:example8} (right panel) shows an example of the
{\sf{}kyhrefl} model spectrum, in which {\sf{}hrefl} spectrum
was smeared across the disk. {\sf{}kyhrefl} can be interpreted
as a Compton-reflection model for which the source of primary 
irradiation is near above the disk, in contrast to the lamp-post
scheme with the source on axis. The approximations for Compton
reflection used in {\sf{}hrefl} (and therefore also in {\sf{}kyhrefl}) 
are valid below $\sim15$~keV in the source (and disk) rest-frame.

\end{document}